\documentclass{emulateapj}
\usepackage{natbib}
\usepackage{graphicx}
\usepackage{bm}

\usepackage[hyperfootnotes=false,naturalnames=true,letterpaper,pdfstartview=FitH,pdfpagemode=none]{hyperref}
\hypersetup{pdfauthor={Dan Coe}, pdftitle={Time Delay Constraints}}

\usepackage{placeins}  

\slugcomment{Accepted for Publication in the Astrophysical Journal}

\newcommand\HO{H_0}
\newcommand\kmsMpc{{\rm km}~{\rm s}^{-1}~{\rm Mpc}^{-1}}
\newcommand\Om{\Omega_m}
\newcommand\OL{\Omega_\Lambda}
\newcommand\Ok{\Omega_k}
\newcommand\Ode{\Omega_{de}}
\newcommand\h{h}
\newcommand\w{{\rm w}}
\newcommand\wo{{\rm w}_0}
\newcommand\wa{{\rm w}_a}

\newcommand\om{\omega_m}

\newcommand\TC{{\mathcal T}_{\mathcal C}}
\newcommand\TCo{{\mathcal T}_{{\mathcal C}\rm ,true}}
\newcommand\TL{{\mathcal T}_{\mathcal L}}
\newcommand\TLiso{{\mathcal T}_{{\mathcal L}\rm ,iso}}
\newcommand\Dt{\Delta \tau}
\newcommand\Dtiso{\Delta \tau_{\rm iso}}
\newcommand\DDt{\Delta (\Delta \tau)}

\newcommand\dDt{\delta (\Delta \tau)}

\newcommand\XC{\Xi_{\mathcal C}}
\newcommand\XL{\Xi_{\mathcal L}}

\newcommand\XLt{\Xi_{{\mathcal L} \tau}}

\newcommand\NL{N_{\rm L}}

\renewcommand\dh{\delta \h}

\newcommand\dTC{\delta \TC}

\newcommand\dXL{\delta \XL}

\newcommand\dXLt{\delta \XLt}

\newcommand\dXi{\delta \Xi}

\newcommand\XZ{\Xi_{\mathcal Z}}
\newcommand\dXZ{\delta \XZ}
\newcommand\XZL{\Xi_{\mathcal Z_L}}
\newcommand\XZS{\Xi_{\mathcal Z_S}}
\newcommand\dXZL{\delta \XZL}
\newcommand\dXZS{\delta \XZS}

\newcommand\Dz{\Delta z}
\newcommand\DzL{\Delta z_L}
\newcommand\DzS{\Delta z_S}

\newcommand\kenv{\kappa_{\rm env}}
\newcommand\klos{\kappa_{\rm los}}

\newcommand\p{\partial}

\newcommand\stat{{\rm stat.}}
\newcommand\syst{{\rm syst.}}

\begin{document}

\title{Cosmological Constraints from Gravitational Lens Time Delays}

\author{
  Dan Coe
  and
  Leonidas A.\ Moustakas
}

\affil{
  Jet Propulsion Laboratory, California Institute of Technology, 
  4800 Oak Grove Dr, MS 169-327, Pasadena, CA 91109 
}

\begin{abstract}
  Future large ensembles of time delay lenses
  have the potential to provide interesting cosmological constraints
  complementary to those of other methods.
  In a flat universe with constant $\w$ including a Planck prior,
  LSST time delay measurements for $\sim 4,000$ lenses
  should constrain the local Hubble constant
  $h$ to $\sim 0.007$ ($\sim 1\%$), 
  $\Ode$ to $\sim 0.005$, and 
  $\w$ to $\sim 0.026$ (all 1-$\sigma$ precisions).
  Similar constraints could be obtained by 
  a dedicated gravitational lens observatory (OMEGA)
  which would obtain precise time delay and mass model measurements
  for $\sim 100$ well-studied lenses.
  We compare these constraints
  (as well as those for a more general cosmology)
  to the ``optimistic Stage IV'' constraints expected from
  weak lensing, supernovae,  baryon acoustic oscillations, and cluster counts,
  as calculated by the Dark Energy Task Force.
  Time delays yield a modest constraint 
  on a time-varying $\w(z)$,
  with the best constraint on $\w(z)$
  at the ``pivot redshift'' of $z \approx 0.31$.
  Our Fisher matrix calculation is provided
  to allow time delay constraints to be easily 
  compared to and combined with 
  constraints from other experiments.
  We also show how cosmological constraining power varies as a function of 
  numbers of lenses, lens model uncertainty, time delay precision, redshift precision,
  and the ratio of four-image to two-image lenses. 
\end{abstract}



\section{Introduction}
\label{sec:intro}

The HST Key Project relied on 40 Cepheids to
constrain Hubble's constant $\HO$
to 11\% \citep{Freedman01}.
The first convincing measurements of the
accelerating expansion rate of the universe
(suggesting the existence of dark energy)
by \cite{Riess98} and \cite{Perlmutter99}
required 50 and 60 supernovae, respectively.
So far, time delays have only been reliably measured for $\sim 16$ gravitational lenses,
thanks to dedicated lens monitoring from campaigns such as COSMOGRAIL \citep{Eigenbrod05}.
Yet recent analyses of 10--16 time delay lenses
already claim to match or surpass the Key Project's 11\% precision on $\HO$
\citep{Saha06a, Oguri07, Coles08}.
Future surveys promise to yield hundreds or even thousands of lenses
with well-measured time delays,
which will enable us to obtain much tighter constraints on $\HO$
as well as constraints on other cosmological parameters.

To date, most efforts have focused on 
studies of individual time delay lenses.
In theory, one might be able to control all systematics and constrain $\HO$ unambiguously 
given a single ``golden lens''.
Such a lens would have a sufficiently simple and well-measured geometry.
The closest to a golden lens may be B1608+656.
In \cite{Suyu09a}, the authors claim all systematics have been controlled to 5\%.
A new estimate for $\HO$ based on this lens is forthcoming \citep{Suyu09b}.

Historically, analyses of individual lenses 
have yielded varying answers for $\HO$
(see the Appendix of \citealt{Jackson07} for a recent review).
This can be attributed to two factors, 
both of which, it appears, are now being overcome.

The first factor is simple intrinsic variation in 
lens properties (especially mass slope) and 
environment (lensing contributions from neighboring galaxies).
Consider the following estimate from a simple empirical argument.
If statistical uncertainties on $\HO$ decrease as $1 / \sqrt{N}$
(assuming systematics can be controlled),
and the current uncertainty from 16 lenses is $\sim 10$\%,
then the uncertainty on a single lens might be $\sim 40$\%.
Thus, assuming $h = 0.7$ (where $\HO = 100 h ~ \kmsMpc$),
individual lenses may be expected to yield a wide range of
$h = 0.42$ -- 0.98 (1-$\sigma$).
(We will revisit these assumptions in this work.)

The second factor in the wide range of reported $\HO$ values
is that different analyses 
have assumed different mass profiles to model the lenses,
including isothermal, de Vaucouleurs, and mass follows light.
There is substantial weight of evidence
that galaxy lenses are roughly isothermal on average,
at least within approximately the scale radius \citep[e.g.,][]{Koopmans06}.
Theoretical work supports this idea,
showing that a wide range of plausible luminous plus dark matter profiles
all combine to yield roughly an isothermal profile {\it at the Einstein radius},
though the slope may deviate from isothermal beyond that radius
\citep{vandeVen09}.

In recent years we have witnessed a steady increase 
in the number of strong lenses discovered by searches such as 
CLASS \citep{CLASS}, 
SLACS \citep{SLACS},
SL2S \citep{SL2S}, 
SQLS \citep{Inada08},
HAGGLeS \citep{Marshall09},
and searches of AEGIS \citep{Moustakas07} and COSMOS \citep{Faure08}.
Based on this experience,
we can expect that future surveys such as
Pan-STARRS\footnote{The Panoramic Survey Telescope \& Rapid Response System, http://pan-starrs.ifa.hawaii.edu} \citep{PanSTARRS}, 
LSST\footnote{The Large Synoptic Survey Telescope, http://www.lsst.org} \citep{LSST}, 
JDEM / IDECS\footnote{The Joint Dark Energy Mission, http://jdem.gsfc.nasa.gov},
and 
SKA\footnote{The Square Kilometer Array, http://www.skatelescope.org} \citep{SKA}
will yield an explosion in
the number of strong lenses known
\citep[e.g.,][]{Koopmans04, Fassnacht04poster, Marshall05}.
Prospects for using these lenses to constrain the nature of dark matter
over the course of the next decade were presented in 
\cite{Moustakas09decadal},  
\cite{Koopmans09decadal}, and
\cite{Marshall09decadal}.

It is reasonable to expect that 
time delays will be reliably measured for large numbers of these lenses, 
whether through repeated observations in surveys (Pan-STARRS and LSST),
auxiliary monitoring, and/or through tailored specific missions 
such as OMEGA \citep{OMEGA}.
Increased sample size, improved lens model constraints,
and higher precision redshifts and time delay measurements
will all improve constraints on $\HO$ and other cosmological parameters,
as we present below.

A more precise measurement of $\HO$ 
will yield tighter constraints on both 
the dark energy equation of state parameter ($\w$) 
and the flatness of our universe ($\Ok$), 
independently of the results of future dark energy surveys
\citep{Blake04,Hu05,DETF06,Olling07}.
To this end, the SHOES Program (Supernovae and $\HO$ for the Equation of State)
has obtained new observations of supernovae and Cepheid variables with reduced systematics.
Recently, \cite{Riess09} published a redetermination of $\HO = 74.2 \pm 3.6 \kmsMpc$,
or 5\% uncertainty including both statistical and systematic errors.
Their $\HO$ determination plus WMAP 5-year data alone
constrain $\w = -1.12 \pm 0.12$ (assuming constant $\w$).

\cite{Riess09} also make the following important point that bears repeating.
The seemingly tight constraints on $\HO$ derived from CMB + BAO + SN experiments
are in fact {\it predictions} or {\it inferences} of $\HO$ 
given those data and a cosmological model.
They are no substitute for direct measurement of $\HO$
such as that presented in their work or the HST Key Project.

\cite{Olling07} reviews several methods with the potential to directly constrain $\HO$.
Water masers, for example, hold much promise \citep{Braatz08, Braatz09}.
Time delays and water masers both yield direct geometric measurements
of the universe to the redshifts of the observed sources
($z \sim 2$ or greater for time delay lenses),
bypassing all distance ladders.

Time delays do not simply constrain $\HO$.
To first order, each time delay is proportional to the
angular diameter distance to the lensed object
and thus inversely proportional to $\HO$.
An additional factor involves a ratio of two other distances --
from observer to lens and from lens to source.
All three of these distances have a complex (though weaker) dependence
on the other cosmological parameters ($\Om, \Ode, \Ok, \wo, \wa$)
which contribute to the expansion history of the universe.

Most time delay analyses ignore this weaker dependence
on ($\Om, \Ode, \Ok, \wo, \wa$),
in effect assuming these parameters are known perfectly.
In this paper we show how relaxing this ``perfect prior''
increases the uncertainties on $\HO$.
As dark energy surveys endeavor to place constraints
on $\w$ and the flatness of our universe $\Ok$,
we must study how time delays
can contribute to these constraints
without assuming the very parameters we would like to constrain.
In this work we also study the ability of large time delay ensembles
to constrain ($\Om, \Ode, \Ok, \wo, \wa$).

The idea to use time delay lenses to measure $\HO$ was 
first proposed by \cite{Refsdal64}.
Strong gravitational lenses are elegant geometric consequences of how
light travels through the universe while grazing massive galaxies.
When the line of sight alignment is very close, 
light takes multiple paths around the curved space of the lens.
These paths form multiple images,
and the light takes a different amount of time to travel each path.
Light passing closer to the lens is deflected by a larger angle
(increasing its path length)
and experiences a greater relativistic time dilation, further delaying its arrival.
If the source flares up, or otherwise varies in intensity
(e.g., if it is an active galactic nucleus, or AGN),
we can observe these ``time delays'' between or among the images.
These time delays are functions of 
the angular diameter distances between the source, lens, and observer,
as well as the properties of the lens itself.

The ability of time delays to constrain other cosmological parameters has also been explored.
\cite{LewisIbata02} explored various combinations of ($\Om, \OL$) in a flat universe
and various ($\wo, \wa$) for fixed ($\Om, \Ode$).
Most notably, they calculated constraints on ($\h, \w$) from ensembles of lenses 
assuming constant $\w$ and ($\Om, \OL$) = (0.3, 0.7),
finding that $\h$ and $\w$ would not be strongly constrained.
We show that the addition of a Planck prior improves these constraints considerably.
\cite{Linder04} investigated constraints on the dark energy parameters ($\wo, \wa$) 
from various methods, touting the complimentarity of strong lensing to that of other methods.
However, they concede that the unique positive correlation 
in strong lensing ($\wo, \wa$) constraints
evaporates when including degeneracies other cosmological parameters.
\cite{Mortsell06} and \cite{Dobke09} examined the constraints 
that large ensembles of lenses might place on $\HO$ and $\OL = 1 - \Om$
(assuming a flat universe).
Below we present the first full treatment of 
the cosmological constraints expected on ($\h, \Om, \Ode, \Ok, \wo, \wa$)
from ensembles of time delay lenses including various priors.

Lens statistics from well-controlled searches for strongly-lensed sources
have also been used to constrain cosmology
\citep[e.g.,][]{Chae07, Oguri08}.
If time delays can be obtained for the lenses in such a sample,
the lens statistics and time delays might combine 
to yield tighter cosmological constraints.
This potential is not explored in this work.

Cosmological constraints can also be obtained from
symmetric strong lenses for which velocity dispersions have been measured
\citep[e.g.,][]{PaczynskiGorski81, FutamaseHamaya99, Yamamoto01, LeeNg07}.
Assuming an isothermal model, 
the measured velocity dispersion determines the Einstein radius
solely as a function of cosmology (given redshifts measured to the lens and source).
\cite{Yamamoto01} studied the future potential for this method
to constrain cosmology using a Fisher matrix analysis.

The reader is invited to skip ahead to our results in \S\ref{sec:cosmofish},
where cosmological constraints expected from time delays (according to our calculations)
are compared to those expected from other methods
(weak lensing, supernovae, baryon acoustic oscillations, and cluster counts).
Table \ref{tab:cosmo} summarizes the assumed priors
including a guide to specific sections and figures.

The remainder of our paper is organized as follows.
In \S\ref{sec:cosmoeq}
we provide the time delay equations
and discuss how cosmology is derived from observed time delays.
We define the quantity $\TC(h, \Om, \Ode, \Ok, \wo, \wa; z_L, z_S)$ 
which time delays are capable of constraining.
In \S\ref{sec:strategy}
we estimate the constraints on $\TC$ expected from future experiments.
(A more detailed analysis of lensing simulations is presented in a companion paper
\citealt[hereafter Paper I]{CoeMoustakas09a}.)
In \S\ref{sec:cosmo}
we illustrate the dependence of $\TC$ on cosmological parameters
($\h, \Om, \Ode, \wo, \wa$).
In \S\ref{sec:cosmofish}, as highlighted above,
we give projections for time delay constraints on $(h, \Ode, \Ok, \wo, \wa)$
and compare to other methods.
Systematic biases are discussed in \S\ref{sec:systematics}
and their impact on our ability to constrain cosmology
is analyzed in another companion paper \cite[hereafter Paper III]{CoeMoustakas09c}.
Finally we present our conclusions in \S\ref{sec:conclusions}.

We assume all constraints to be centered on the concordance cosmology
$h = 0.7$, $\Om = 0.3$, $\Ode = 0.7$, $\Ok = 0$, $\wo = -1$, and $\wa = 0$,
where $\HO = 100 h ~ \kmsMpc$.

\section{Cosmological Constraints from Time Delays}
\label{sec:cosmoeq}

\subsection{Time Delay Equations}
\label{sec:lenseq}

A galaxy at redshift $z_L$
strongly lenses a background galaxy at redshift $z_S$
to produce multiple images.
Either two or four images are typically produced.\footnote{An additional
central demagnified image is also produced by
every lens with a central mass profile shallower than isothermal.
Such images are rarely bright enough to be detected,
thus we ignore them throughout this work.}
We refer to these cases as ``doubles'' and ``quads'', respectively.
The lensing effect delays each image in reaching our telescope
by a different amount of time, given by
\\
\begin{equation}  %
  \label{eq:timedelay1}
  \Delta \tau = \frac{(1 + z_L)}{c} {\mathcal D}
  \left [
  \onehalf \left \vert {\bm \theta} - {\bm \beta} \right \vert ^2 - \phi
  \right ]
\end{equation}
\\
\citep[e.g.,][]{BlandfordNarayan86}
with terms defined below.  
The factors in the time delay equation can be grouped into a product of two terms:
\\
\begin{equation}  %
  \label{eq:dtCL}
  \Delta \tau = \TC \TL.
\end{equation}

The first factor,
\\
\begin{equation}  %
  \label{eq:TC}
  \TC \equiv \frac{(1 + z_L)}{c} {\mathcal D},
\end{equation}
\\
is a function of cosmology 
and the lens and source redshifts, $z_L$ and $z_S$.
The second factor,
\\
\begin{equation}  %
  \label{eq:TL}
  \TL \equiv
  \left [
    \onehalf \left \vert {\bm \theta} - {\bm \beta} \right \vert ^2 - \phi
  \right ],
\end{equation}
\\
is a function of the projected lens potential $\phi$,
the source galaxy's position on the sky $\bm \beta$,
and the image positions $\bm \theta$.

We concentrate on the cosmological dependence of $\TC$.
The factor
\\
\begin{equation}  %
  {\mathcal D} \equiv \frac{D_L D_S}{D_{LS}}
\end{equation}
\\
is a ratio of the angular-diameter distances from
observer to lens $D_L = D_A(0,z_L)$,
observer to source $D_S = D_A(0,z_S)$,
and lens to source $D_{LS} = D_A(z_L,z_S)$.
Angular-diameter distances are calculated as follows 
\citep[filled beam approximation; see also \citealt{Hoggcosmo}]{Fukugita92}:
\\
\begin{equation}
  D_A(z_1,z_2) = \frac{c}{\HO} \frac{E_A(z_1,z_2)}{1 + z_2},
\end{equation}

\begin{equation}
  \label{eq:EAcurved}
  E_A = \frac{{\rm sinn}
    \left[ \sqrt{\left \vert \Ok \right \vert} E^\star_{A} \right]}
    {\sqrt{\left \vert \Ok \right \vert}},
\end{equation}
\\
where ${\rm sinn}(u) = \sin(u)$, $u$, or $\sinh(u)$
for an open, flat, or closed universe respectively
($\Ok < 0$, $\Ok = 0$, or $\Ok > 0$).
The curvature is given by $\Ok \equiv 1 - (\Om + \OL)$, while
\\
\begin{equation}
  \label{eq:EA}
  E^\star_{A}(z_1,z_2) = \int_{z_1}^{z_2}
  \frac{dz^\prime}{E(z^\prime)}.
\end{equation}

The normalized Hubble parameter $E(z)$
can have different expressions
depending on the cosmology assumed:
\\
\begin{eqnarray}
  \label{eq:E}
  E(z) & \equiv & \frac{H(z)}{\HO}\\
  & = & \sqrt{
    \Om (1 + z)^3 +
    \Ok (1 + z)^2 +
    \OL} \nonumber \\
  & = & \sqrt{
    \Om (1 + z)^3 +
    \Ok (1 + z)^2 +
    \Ode (1 + z)^{3 (1 + \w)}} \nonumber \\
  & = & \sqrt{
    \cdots +
    \Ode (1 + z)^{3 (1 + \wo + \wa)} 
    \exp{ \left( \frac{-3 \wa z}{1 + z} \right) }} \nonumber.
\end{eqnarray}
\\
Here we have progressed from a universe with a cosmological constant $\OL$\ to
one with dark energy with an equation of state $p = \w \rho$.
In the last line, the last term has been rewritten 
in terms of an evolving dark energy equation of state
\\
\begin{eqnarray}
  \label{eq:w}
  \w & = & \wo + \wa (1 - a) \\
       & = & \wo + \wa \left( \frac{z}{1 + z} \right),
\end{eqnarray}
\\
a common parametrization first introduced by 
\cite{ChevallierPolarski01} and \cite{Linder03}.
The universe scale factor $a = (1 + z)^{-1}$.

We next define the dimensionless ratio
\\
\begin{equation}  %
  {\mathcal E} \equiv \frac{E_L E_S}{E_{LS}}
\end{equation}
\\
with factors defined similarly to those above for $D_A$:
$E_L = E_A(0,z_L)$,
$E_S = E_A(0,z_S)$,
$E_{LS} = E_A(z_L,z_S)$.
We find that many factors cancel, and $\TC$ simplifies to:
\\
\begin{equation}  %
  \label{eq:TCEHO}
  \TC = \frac{\mathcal E (\Om, \Ode, \Ok, \wo, \wa)}{\HO}.
\end{equation}
\\
We see here clearly that time delays 
($\Dt = \TC \TL$)
scale inversely with $\HO$.
There is also a complex though weaker dependence 
on the other cosmological parameters
as embedded in $\mathcal E$.

\subsection{Deriving Cosmology from Time Delays}

Given observed time delays $\Dt$
and assuming a lens model (and thus $\TL$),
one can obtain measures of $\TC$.
These measures will have some scatter
due to both observational uncertainties
and deviations of the lens from the assumed model.

Recent studies suggest that galaxy lenses, on average,
have roughly isothermal profiles within the Einstein radius
(see \S \ref{sec:intro}).
Deviations from this simple description include
variation in lens slope,
external shear,
mass sheets,
and substructure.
\cite{Oguri07} parametrized the deviations as
the ``reduced time delay'', the ratio of the observed time delay
to that expected due to an isothermal potential in a given lens:
\\
\begin{equation}  %
  \label{eq:dtred}
  \Xi \equiv \frac{\Dt}{\Dtiso}.
\end{equation}
\\
In our notation, these observed deviations
are due to deviations in the lens model:
\\
\begin{equation}  %
  \label{eq:XL}
  \XL \equiv \frac{\TL}{\TLiso}.
\end{equation}
\\
By assuming an isothermal model ($\TL = \TLiso$),
these deviations get absorbed into the derived cosmology:
\\
\begin{equation}  %
  \label{eq:XC}
  \XC \equiv \frac{\TC}{\TCo},
\end{equation}
\\
where $\TCo$ is the true cosmology.
For example, a lens which is steeper than isothermal yields $\XL > 1$;
thus when assuming an isothermal model ($\XL = 1$),
we derive $\XC > 1$
(since $\Xi = \XC \XL$).
In traditional analyses assuming fixed $\mathcal E$,
$\XC > 1$ would simply yield a low $\h$.
This approximation is adequate for small samples of lenses
but not for the large samples to come in the near future (\S\ref{sec:relax}).

Similarly, observational uncertainties affecting $\Dt$
are absorbed into the derived cosmology.
In this paper, 
we study how observational and intrinsic (lens model) uncertainties
combine to yield scatter in the observed $\Dt$.
We will assume these measurements yield $\TC$
with the correct mean but a simple Gaussian scatter
and explore how this propagates to Gaussian uncertainties on 
cosmological parameters.

In practice we do not expect
$\XL$ and measurements of $\Dt$ to have Gaussian scatter,
but these serve as useful approximations.
The true expected $P(\Xi)$ from time delay measurements 
and methods for handling these distributions
are studied in \cite{Oguri07} and Paper I.

\section{Constraints on $\TC$ from Future Experiments}
\label{sec:strategy}

\subsection{Extrapolating from Current Empirical Results}
\label{sec:extrap}

Recent studies have constrained $\TC$ to $\sim 10\%$ using time delays,
where $\TC$ encodes all of the cosmological dependencies (\S\ref{sec:lenseq}).
Constraints on $\TC$ have generally been 
interpreted to be equivalent to direct constraints on $h$.
This assumption is reasonable for current sample sizes,
but will need to be revised in the future (\S\ref{sec:relax}).
Using 16 lenses,
\cite{Oguri07} obtain 
$\h = 0.70 \pm 0.06{\rm (stat.)}$.
Similar studies by \cite{Saha06a} and \cite{Coles08}
using a different method
obtain similar constraints using 10 and 11 lenses, respectively.
The latter finds $\h = 0.71^{+0.06}_{-0.08}$.

We will adopt the \cite{Oguri07} uncertainty of 8.6\% with 16 lenses
as the ``current'' uncertainty in $\TC$.\footnote{The \cite{Oguri07} simulations
initially suggested an uncertainty of $\sim 4\%$ in $\TC$.
However jackknife resampling of the data revealed the true uncertainty to be twice as much.
Under-prescribed shear in the simulations was cited as a potential cause for the discrepancy.}
We note that the time delay uncertainties in this sample 
are roughly and broadly scattered about $\DDt = 2$ days.\footnote{We adopt 
a notation in which ``$\Delta$'' refers to uncertainties with units
and ``$\delta$'' to fractional uncertainties.
Thus a time delay of 20 days measured to 2-day precision has
$\DDt = 2$ days and $\dDt = 0.1$.}

We can improve on these $\TC$ constraints in three ways:
obtaining larger samples of lenses,
better constraining our lens models,
and obtaining more precise time delay measurements.
As we explain below, we expect future surveys such as Pan-STARRS and LSST
to improve on the sample size
while the lens model and time delay uncertainties will remain about the same.
These surveys will have to contend with a lack of spectroscopic redshifts for most objects,
but the gains in sample size will more than compensate.
Similarly tight constraints on $\TC$ could also be obtained
by studying relatively fewer lenses in great detail, as we discuss below.

Here we consider statistical uncertainties only,
with systematics to be discussed in \S\ref{sec:systematics}.
We will assume that 
all other things being equal,
increasing our sample size beats down our errors by 
$\sqrt{N}$ for $N$ lenses.
This assumption is borne out well by our detailed simulations (Paper I),
for the case of no systematic uncertainties.

Based on the current constraint of $\dTC \approx 8.6\%$ from 16 lenses \citep{Oguri07},
we project that simply increasing the sample of lenses
would produce constraints of $\dTC \approx 34\% / \sqrt{N}$.
We will define this as the uncertainty from lens models and time delay measurements:
$\dXLt \sim 0.344$.
Photometric redshifts would degrade these constraints as estimated below (\S\ref{sec:dz}).

\subsection{Future Surveys}
\label{sec:exp}

Pan-STARRS and LSST will both survey the sky repeatedly,
opening the time domain window for astronomical study over vast solid angles.
Pan-STARRS 1 (PS1) has recently begun its $3\pi$ survey,
repeatedly observing the entire visible sky to $\sim$23rd magnitude every week
over a 3-year period.
LSST promises similar coverage and depth every 3 nights
with first light scheduled for 2014.

These surveys will reveal many time-variable sources,
among them gravitationally-lensed quasars.
The persistent monitoring over many years
should yield time delays ``for free'' for many strongly-lensed quasars.
Simulations (M.~Oguri 2009, private communication)
show that Pan-STARRS 1 and LSST are expected to yield
$\sim 1,000$ and $\sim 4,000$ strongly-lensed quasars
with quad fractions of 19\% and 14\%, respectively.

We will assume that these surveys will measure time delays to about 2-day precision,
or similar to that of our current sample of time delay lenses.
This is consistent with predictions based on detailed simulations by \cite{Eigenbrod05}
which study factors including survey cadence, object visibility,
and the complicating effects of microlensing.
We note this estimate may be a bit optimistic for PS1
with its slower sampling rate compared to LSST.

\begin{figure}
\plotone{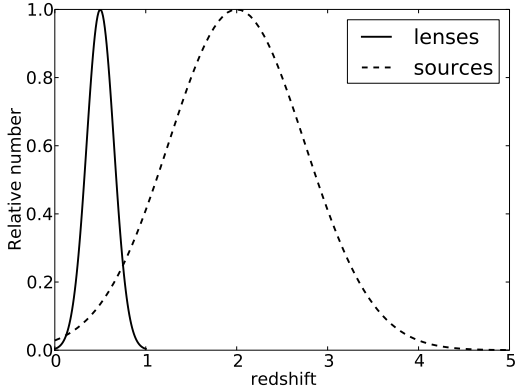}
\\
\caption[zdist]{
\label{fig:zdist}
Distributions of lens and source redshifts used in this paper.
These Gaussian distributions
($z_L = 0.5 \pm 0.15$,
$z_S = 2.0 \pm 0.75$;
$z_S > z_L$)
were used by \cite{Dobke09}
as reasonable approximations for near-future missions including LSST.
}
\end{figure}

The expected redshift distributions of the lenses and sources
can be roughly approximated by the Gaussian distributions
$z_L = 0.5 \pm 0.15$ and
$z_S = 2.0 \pm 0.75$
with $z_S > z_L$
(Fig.~\ref{fig:zdist}), as adopted by \cite{Dobke09}.
Obviously the two distributions will be correlated, but we approximate them as being independent.

As surveys attain fainter magnitude limits,
it is believed that the magnification bias enjoyed by quads will be diminished.
Future surveys are thus expected to yield lower quad fractions ($\sim 19\%, 14\%$)
than the current sample of time delay lenses (6 / 16 = 37.5\%).
This might improve the expected constraints on $\TC$ from future surveys
as quads have been shown to yield time delays with more scatter
and thus less reliable estimates of $\TC$ \citep[Paper I]{Oguri07}.\footnote{This is
believed to be due to the fact that some of the factors (especially external shear)
which cause scatter in $\Xi$ 
also raise the likelihood that a lens will produce quad images rather than a double.}
However, we find this to be mitigated by the fact that 
quads yield multiple time delay measurements (one for each pair of images),
while doubles only yield a single $\Dt$ measurement.
Based on our detailed simulations and analysis (Paper I),
we find quads and doubles to have approximately equal power to constrain $\TC$.
This simplifies our analysis;
the quad-to-double ratio need not be considered
when estimating $\dTC$ for a given experiment.
To allay any concern, 
we stress that this assumption actually makes our estimates of $\dTC$ more conservative 
for future surveys which have lower quad fractions than the current sample.

For each double or quad, image pairs can be further classified by their geometry.
For example, image pairs with small opening angles are found to yield larger scatter in $\Dt$
\citep[Paper I]{Oguri07}.
Detailed analyses in these papers quantify these scatters,
enabling a well-informed prior $P(\Xi)$ to be placed on each image pair 
as a function of geometry.
The details are unimportant here 
though we have made use of the constraint this analysis has put on $\TC$ \citep{Oguri07}.

\subsection{Photometric Redshift Uncertainties}
\label{sec:dz}

Currently all lenses which have reliable time delay measurements
also have spectroscopic redshifts measured for both lenses and sources
\citep[e.g.,][]{Oguri07}.
The telescope time required to obtain spectroscopic redshifts
is generally a small fraction of that required to obtain accurate time delays,
so the extra investment is worthwhile.

Future surveys which repeatedly scan the sky, however,
will yield time delays for many more lenses 
than may be followed up spectroscopically.
For these lenses we will have to rely on photometric redshift measurements.
These uncertainties will degrade the constraints possible on the cosmological parameters.

Photometric redshift uncertainties for the lenses
(typically elliptical galaxies at $z_L \sim 0.5$)
are expected to be $\DzL \sim 0.04 (1 + z_L)$,
similar to that found in the CFHT Legacy Survey \citep{Ilbert06}.
Redshift uncertainties for the lensed sources (quasars)
are expected to be somewhat higher.
We will adopt $\DzS \sim 0.10 (1 + z_S)$,
roughly that found in the analysis of $\sim$one million SDSS quasars \citep{Richards09}.

Obtaining photometric redshifts in ground-based images
will often be complicated by 
cross-contamination of flux among the lens and multiple images.
Yet improved photometric redshift techniques are also being developed with LSST in mind
\citep{Schmidt09AAS},
so it is perhaps too early to say whether our estimated redshift uncertainties
are too optimistic or pessimistic for a future ground-based survey.
Some of the most common catastrophic redshift degeneracies can clearly be avoided
by considering the observed image separations, time delays, etc.
Most obviously, 
the common degeneracy between $z \sim 0.2$ and $z \sim 3$ \citep[e.g.,][]{Coe06} 
can be neatly averted since a lens at $z \sim 3$ or a source at $z \sim 0.2$
would clearly stand out.

Assuming the above redshift uncertainties, 
we now determine how these propagate into uncertainties on $\TC$.
For simplicity, let us assume that redshift uncertainties are Gaussian.
Let us further assume that uncertainty in $\Xi$ scales linearly with redshift uncertainty.
(This is approximately true for reasonable uncertainty levels $\Delta z \lesssim 0.2$.)

\begin{figure*}
\plottwo{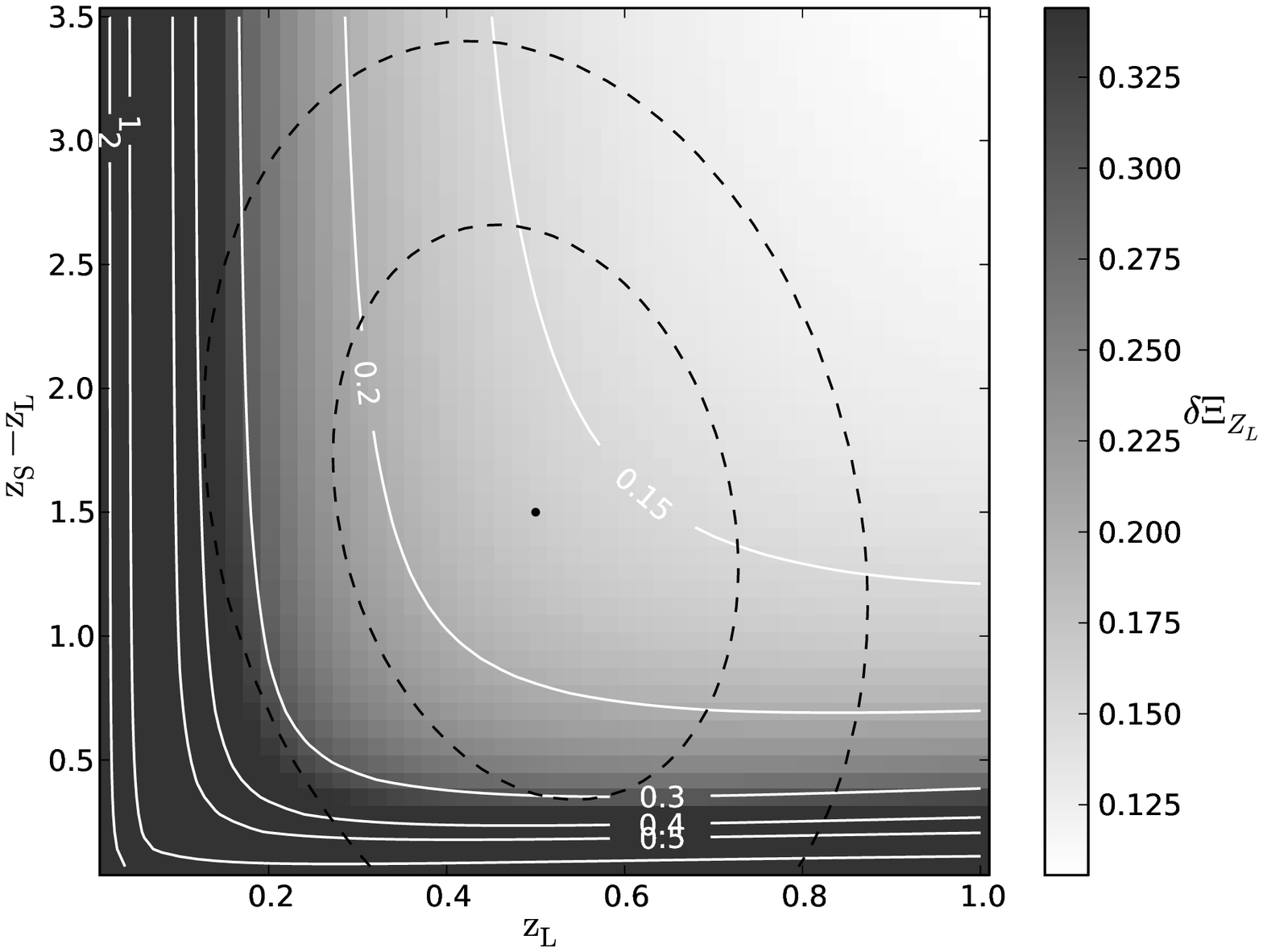}{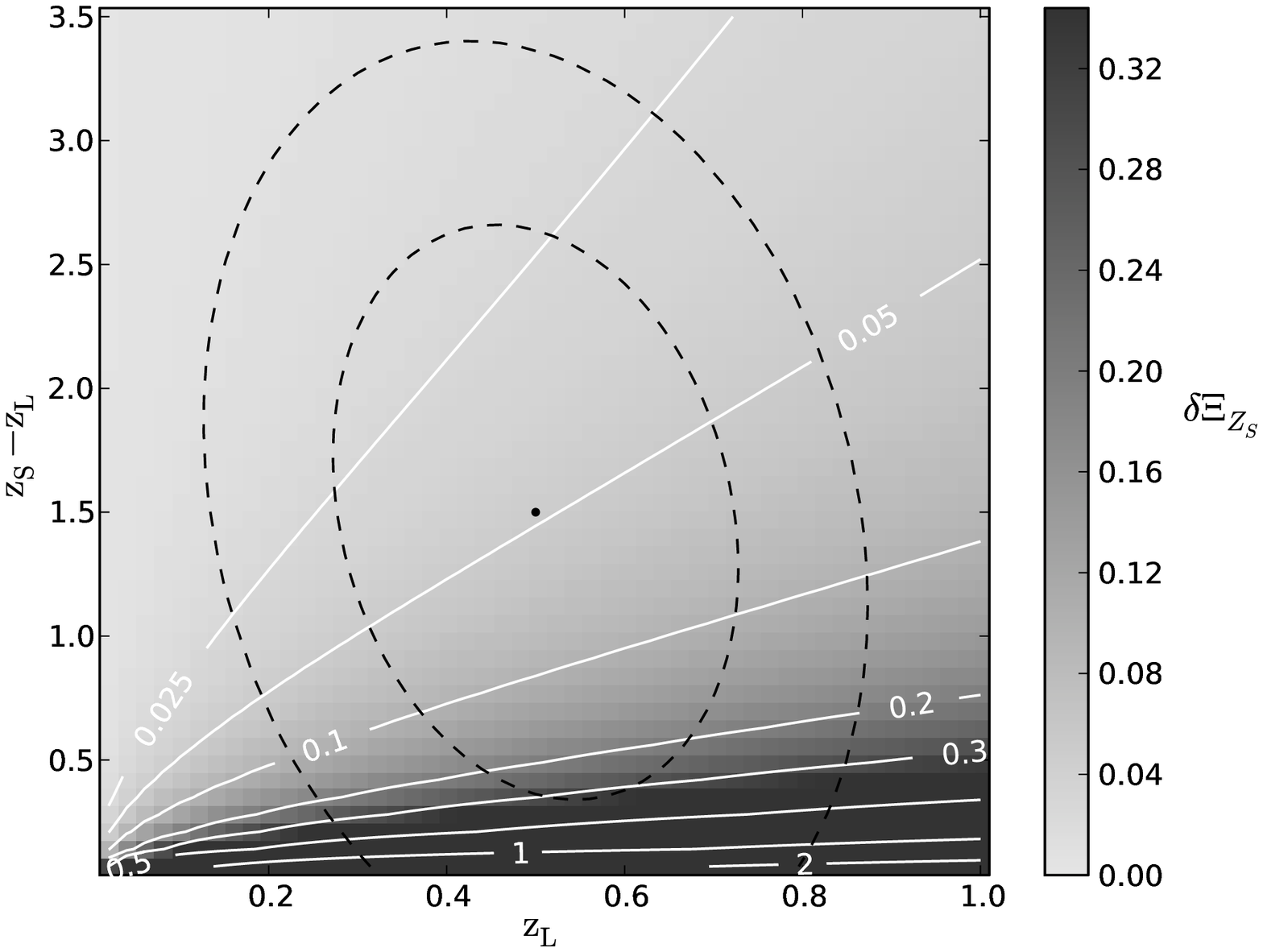}
\\
\caption[dtdzs]{
\label{fig:dtdzs}
Photometric redshift uncertainties'
contributions to cosmological uncertainties in $\TC$.
{\it Left}: Uncertainty in $\TC$ (grayscale and contours) 
from lens redshift uncertainties of $0.04 (1 + z_L)$,
plotted as a function of lens redshift $z_L$
and the lens-source redshift difference $z_S - z_L$.
The dashed contours show the redshift distribution (1- and 2-$\sigma$ contours) 
assumed in this work.
A dot at $(z_L, z_S) = (0.5, 2.0)$ marks the center of the distributions.
{\it Right}: Same for source redshift uncertainties of $0.10 (1 + z_L)$.
Note that the plots have different grayscales.
For sources close to the lens (small $z_S - z_L$),
redshift uncertainties become catastrophic yielding large $\dTC$.
Lens redshift uncertainties are also problematic at low $z_L$.
}
\end{figure*}

Using equations \ref{eq:EAcurved} -- \ref{eq:TCEHO},
we find for a typical lens-source combination with $(z_L, z_S) = (0.5, 2.0)$,
that lens and source redshift uncertainties translate to
$\dXZL \sim 2.75 \DzL$ and
$\dXZS \sim -0.16 \DzS$, respectively.
Given the above redshift uncertainties, these evaluate to
$\dXZL \sim 0.16$ and
$\dXZS \sim 0.05$.
These relations are strong functions of redshift
and become catastrophic for sources very close to the lens.
We plot this behavior in Fig.~\ref{fig:dtdzs}.
If accurate and precise redshifts are not available,
we must concentrate our analysis on systems with high separation in redshift
between the lens and source.

For a lens ensemble with Gaussian redshift distributions
$z_L = 0.5 \pm 0.15$ and
$z_S = 2.0 \pm 0.75$,
we find
$\dXZL \sim 0.175$ and
$\dXZS \sim 0.028$.
To calculate these uncertainties,
we sum the $\chi^2$ of individual lens-source combinations,
weighting by the probability $P_i$ of observing that combination:
\\
\begin{equation}  %
  \label{eq:invquad}
  \frac{1}{\sigma^2} = \sum_i \frac{P_i}{\sigma_i^2}.
\end{equation}
\\
Note that this sum naturally assigns more weight to more confident measurements.

Assuming the lens and source redshift uncertainties can be added in quadrature,
\\
\begin{equation}  %
  \dXZ^2 = \dXZL^2 + \dXZS^2,
\end{equation}
\\
we find $\dXZ \sim 0.177$.

Of course, these are just estimates for large ensembles.
In practice, redshift probability distributions $P(z)$ for individual galaxies
will be properly folded into the $P(\TC)$ determinations.
Biased redshifts would yield biased $\TC$,
the effects of which we study in Paper III.

\subsection{Projected Constraints from Large Surveys}

We now calculate the total uncertainty $\dTC$ expected for large surveys
with photometric redshifts.
The combined lens model and time delay uncertainties are $\dXLt \sim 0.344$,
based on extrapolation of the current empirical \cite{Oguri07} finding (\S\ref{sec:extrap}).
We estimate uncertainties of $\dXZ \sim 0.177$
due to redshift uncertainties of 
$\DzL \sim 0.04 (1 + z_L)$ and
$\DzS \sim 0.10 (1 + z_S)$
for the lenses and sources, respectively (\S\ref{sec:dz}).

The simplest estimate of the total uncertainty 
is to add these uncertainties in quadrature:
\\
\begin{equation}  %
  \dXi^2 = \dXLt^2 + \dXZ^2.
\end{equation}
\\
This yields $\dXi \sim 0.387$.

To be more precise, all of the uncertainties should be added in quadrature
for each lens individually
before combining them according to Eq.~\ref{eq:invquad}.
Repeating the analysis in this way, we find $\dXi \sim 0.402$.

Thus we expect large surveys with photometric uncertainties given above
to yield $\dTC \sim 40\% / \sqrt{N}$.
We project $\dTC \sim 1.3\%$ for PS1 (1,000 lenses)  
and $\dTC \sim 0.64\%$ for LSST (4,000 lenses).  

Table \ref{tab:dTCexp} summarizes the progress we can expect to make
in ``Stages'' corresponding to those defined by the 
Dark Energy Task Force (DETF; \citealt{DETF06,DETF09}):
``Stage I'' = current, ``II'' = ongoing, ``III'' = currently proposed, ``IV'' = large new mission.
Again, we stress these are estimates of statistical uncertainties only.
Large surveys are compared to 
dedicated monitoring and detailed analysis of a smaller sample of lenses.

We might have made our analysis more sophisticated still,
calculating $\dXLt$, $\dXZL$, and $\dXZS$ individually
for each lens-source combination in our ensemble.
Lenses and sources at higher redshift, for example,
will be brighter and higher magnification cases on average,
altering their $\dXLt$ somewhat.
The approximations made in our above analysis should suffice for our purposes here.

\begin{deluxetable*}{clrccccl}
\tablecaption{Estimated Current and Future Constraints on $\TC$\label{tab:dTCexp}}
\tablehead{
\colhead{Stage} &
\colhead{Experiment} &
\colhead{$\NL$} &
\colhead{quads} &
\colhead{$\Dz$\tablenotemark{a}} &
\colhead{$\DDt$} &
\colhead{$\dXL$} &
\colhead{$\dTC$}
} 
\startdata
I & current & 16 & 38\% & spec & 2 days & \nodata & 8.6\%\\
II & Pan-STARRS 1 & 1,000 & 19\% & phot & 2 days & \nodata & 1.27\%\\  
IV & LSST & 4,000 & 14\% & phot & 2 days & \nodata & 0.64\%\\
IV & OMEGA & 100 & 100\% & spec & 0.1 day & 5\% & 0.5\%\\
IV & LSST + OMEGA & \nodata & \nodata & \nodata & \nodata & \nodata & 0.4\%\\
\vspace{-0.1in}
\enddata
\tablenotetext{a}{Spectroscopic or photometric redshift measurements.
For the latter we assume $\DzL = 0.04(1+z_L)$ and $\DzS = 0.10(1+z_S)$.}
\end{deluxetable*}

\begin{figure}
\includegraphics[width=\hsize]{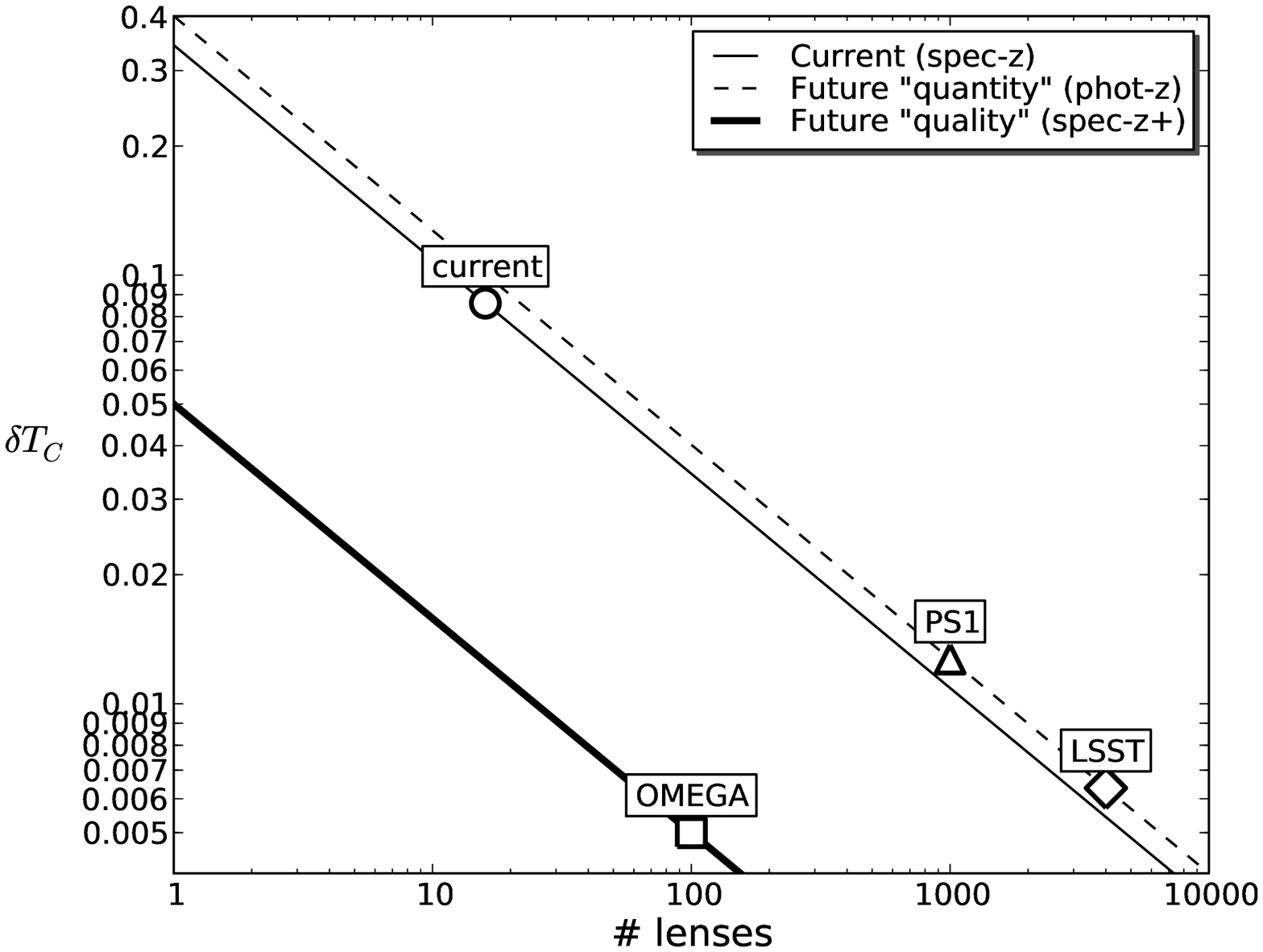}
\caption[TCNall]{
\label{fig:TCNall}
Constraints on $\dTC$ as a function of ensemble size
and observational uncertainties.
The current ensemble has time delays measured to roughly $\DDt = 2$ day precision
and spectroscopic redshifts measured for all lenses and sources.
Future large surveys (``quantity'') should have similar time delay precisions
but photometric redshifts measured for 
lenses ($\DzL = 0.04(1+z_L)$) and
sources ($\DzS = 0.10(1+z_S)$).
A dedicated campaign (``quality'') could in principle obtain 
tight lens model constraints ($\dXL = 5\%$)
with high-precision time delays ($\DDt = 0.1$ day)
and spectroscopic redshifts.
}
\end{figure}

\subsection{Quality vs.~Quantity}
\label{sec:quality}

Thus far we have assumed that detailed observations and analysis
would not be performed on the lenses.
The alternative is to study fewer lenses in more detail,
reducing the uncertainties for each lens.
In practice, we expect both strategies to be pursued
and the combined power of both analyses to place the tightest possible constraints on $\TC$.

\cite{OMEGA} have designed a mission concept
that would be dedicated to monitoring a sample of four-image lenses,
with the primary goal of constraining fundamental properties of dark matter.
This space-based Observatory for Multi-Epoch Gravitational Lens Astrophysics (OMEGA)
would monitor 100 time delay lenses
to achieve precise and accurate $\lesssim 0.1$ day time delay measurements. 
Supporting measurements would aim to reduce the model uncertainty of each lens to 5\%
($\dXL = 0.05$)
and thus constrain $\TC$ to 5\% with each lens,
as claimed recently for B1608+656 \citep{Suyu09b}.
These supporting measurements, including velocity dispersion in the lens
and characterization of the group environment (see discussion in \S\ref{sec:masssheet}),
would be carried out either with OMEGA itself
or though coordinated efforts by ground-based telescopes and JWST.
Spectroscopic redshifts would also be obtained for the 100 lens galaxies and lensed quasars.

Lenses targeted by OMEGA will be quads,
enabling measurements of time delay \textit{ratios} among the image pairs.
This would provide constraints on the dark matter substructure mass function
\citep[Moustakas et al., in preparation]{KeetonMoustakas09, Keeton09}.

Given lens models accurate to 5\% for 100 galaxies,
we might expect OMEGA to yield $\dTC \sim 5\% / \sqrt{100} = 0.5\%$.
The time delays would be measured with sufficient precision
so as not to contribute significantly to the total uncertainty in $\dTC$.
The multiple time delay measurements per lens (quad)
also help reduce this contribution.
Based on the expected time delay distribution for a sample of quads (Paper I),
we estimate that $\DDt = 0.1$-day uncertainties 
would inflate the $\TC$ uncertainty only to $\sim0.515\%$.

If both LSST and OMEGA obtain their measurements of $\TC$ free of significant systematics,
their combined power could further reduce the uncertainty to $\dTC \sim 0.4\%$.

\section{Dependence of $\TC$ on Cosmology}
\label{sec:cosmo}

We expect LSST time delay lenses to constrain $\TC$ to $\sim 0.64\%$.
In this section we begin to explore how this ``Stage IV'' constraint
translates to constraints on cosmological parameters.
We study the dependence of $\TC$ on $(h, \Om, \Ode, \Ok, \wo, \wa)$
for several cosmologies as outlined in Table \ref{tab:cosmo}.

\begin{deluxetable*}{lllllllll}
\tablecaption{Cosmologies explored in this work\label{tab:cosmo}}
\tablewidth{\textwidth}
\tablehead{
\colhead{Cosmology} &
\colhead{$h$} &
\colhead{$\Om$} &
\colhead{$\Ode$ / $\OL$\tablenotemark{a}} &
\colhead{$\Ok$} &
\colhead{$\wo$} &
\colhead{$\wa$} &
\colhead{Sections\tablenotemark{b}} &
\colhead{Figures}
} 
\startdata
Flat universe with cosmological constant & Free & $1 - \OL$ & Free ($\OL$) & 0 & $-1$ & 0 & \S\ref{sec:flatcosmoconst} 
& \ref{fig:hOLflat}, \ref{fig:hOLflat_ensemble}\\
Curved universe with cosmological constant & Free & $1 - (\OL + \Ok)$ & Free ($\OL$) & Free & $-1$ & 0 & \S\ref{sec:curvedcosmoconst} 
& \ref{fig:OmOLh}, \ref{fig:OmOLh_ensemble}\\
Flat universe with constant $\w$\tablenotemark{c} & Free & $1 - \Ode$ & Free & 0 & Free & 0 & \S\ref{sec:flatwconst}, \S\ref{sec:Fisherflatwconst} 
& \ref{fig:Hwflat}, \ref{fig:Hwflat_ensemble}, \ref{fig:StageIVflatwconst}\\
Flat universe with time-variable $\w$ & Free & $1 - \Ode$ & Free & 0 & Free & Free & \S\ref{sec:flatwvar} 
& \ref{fig:w0wah}\\
General (curved with time-variable $\w$)\tablenotemark{d} & Free & $1 - (\Ode + \Ok)$ & Free & Free & Free & Free & \S\ref{sec:cosmogen} 
& \ref{fig:FishPlanck}, \ref{fig:StageIV}, \ref{fig:tdvh}\\
\vspace{-0.1in}
\enddata
\tablecomments{We consider six cosmological parameters
of which five are independent since $\Om + \Ode + \Ok = 1$.
}
\tablenotetext{a}{When $\wo = -1$ and $\wa = 0$, $\Ode = \OL$, the cosmological constant.}
\tablenotetext{b}{In \S\ref{sec:cosmo} the $\TC$ dependencies are explored.
In \S\ref{sec:cosmofish} additional priors are assumed and time delay constraints are compared to those from other methods.}
\tablenotetext{c}{Given this cosmology, we assume a Planck prior in \S\ref{sec:Fisherflatwconst}.}
\tablenotetext{d}{Given a general cosmology, in \S\ref{sec:cosmogen}
we assume a prior of Planck + ``Stage II'' WL+SN+CL
(see that section for details).}
\end{deluxetable*}

\subsection{Flat universe with a cosmological constant ($h$, $\OL = 1 - \Om$)}
\label{sec:flatcosmoconst}

\begin{figure}
\plotone{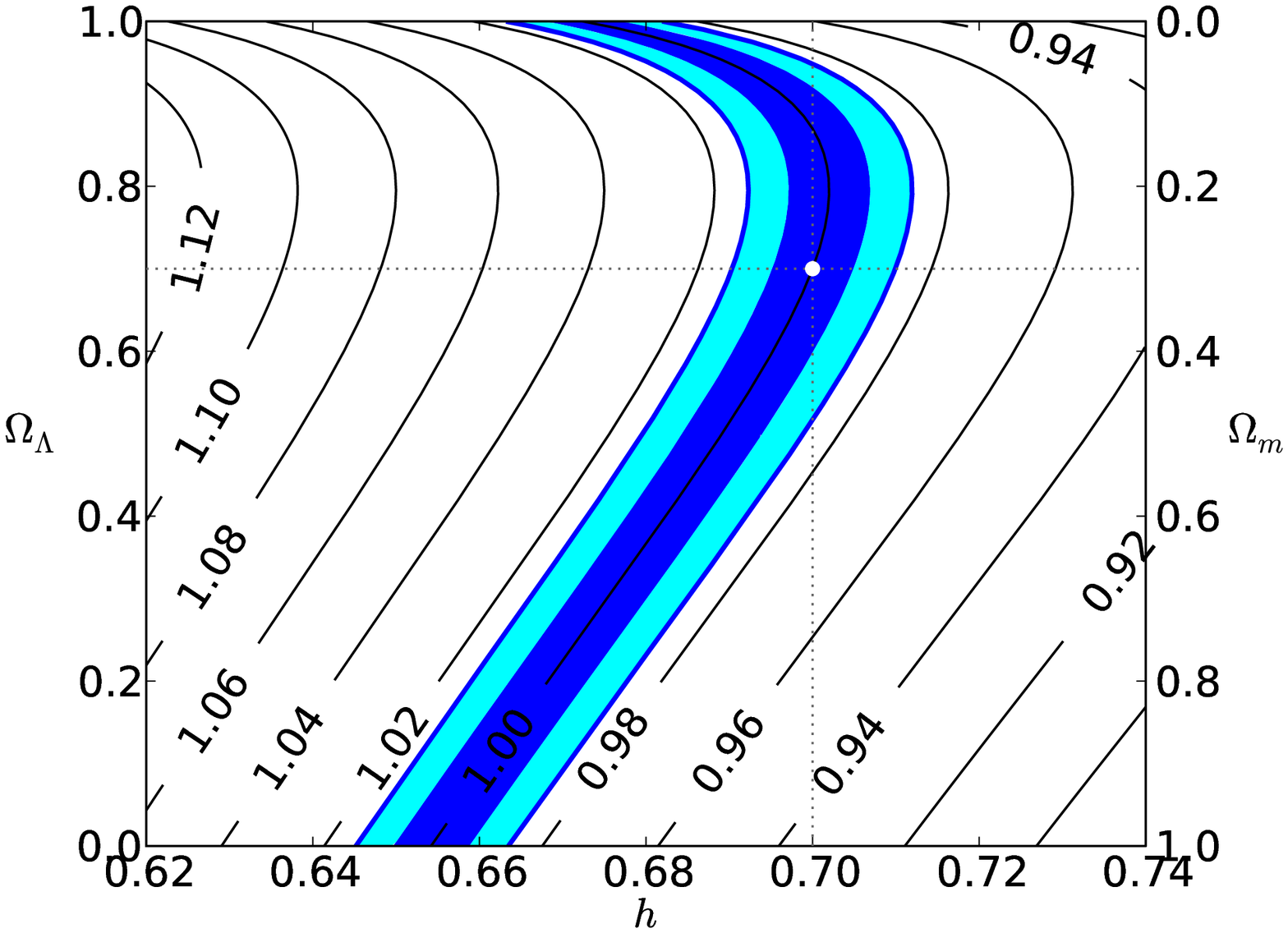}
\\
\caption[h0L]{
\label{fig:hOLflat}
Confidence contours (1- and 2-$\sigma$ colored bands) for
($h$, $\OL = 1 - \Om$)  
given ``Stage IV'' $\dTC = 0.64\%$ obtained from an ensemble with
all lenses and sources at $z_L$, $z_S$ = (0.5, 2.0).
Here we assume a flat universe with a cosmological constant ($\w = -1$).
Also plotted are contours of constant $\XC \equiv \TC / \TCo$,
where $\TCo \approx 0.99$ for the input redshifts and cosmology.
The input cosmology ($h, \Om, \OL$) = (0.7, 0.3, 0.7)
is marked with dotted lines and a white dot.
}
\end{figure}

First, we add a single free parameter $\OL$ (in addition to $h$)
in considering a flat universe with a cosmological constant ($\w = -1$).
Given $\dTC = 0.64\%$ from an ensemble
with {\it all} lenses at $z_L = 0.5$ and {\it all} sources at $z_S = 2.0$,
we would obtain confidence contours shown in Fig.~\ref{fig:hOLflat}.

\begin{figure}
\plotone{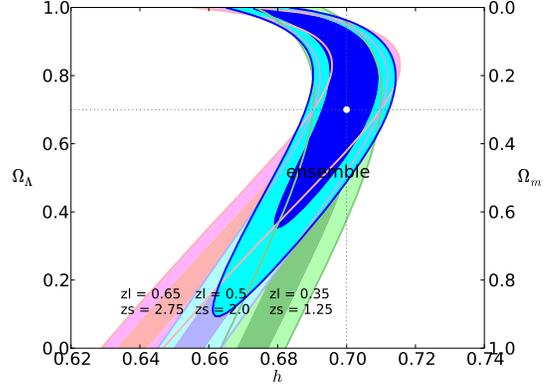}
\\
\caption[h0Lflat]{
\label{fig:hOLflat_ensemble}
Confidence contours (1- and 2-$\sigma$ colored bands) for
($h$, $\OL = 1 - \Om$)  
given $\dTC = 0.64\%$
and assuming a flat universe with a cosmological constant ($\w = -1$).
Each of the three fainter curves corresponds to
all lenses and sources at the same pair of redshifts:
$z_L$, $z_S$ = (0.65, 2.75), (0.5, 2.0), (0.35, 1.25), as marked. 
Next we consider an ensemble 
of lenses and sources with Gaussian redshift distributions:
$z_L$, $z_S$ = ($0.5 \pm 0.15$, $2.0 \pm 0.75$).
These yield the tighter constraints (marked ``ensemble'').
The input cosmology ($h, \Om, \OL$) = (0.7, 0.3, 0.7)
is marked with a white dot.
}
\end{figure}

The shape of these curves shifts somewhat as a function of $z_L$ and $z_S$.
Given an ensemble of lenses and sources 
with Gaussian redshift distributions
$z_L = 0.5 \pm 0.15$ and $z_S = 2.0 \pm 0.75$ as discussed above,
we begin to break the ($h, \OL$) degeneracy (Table \ref{fig:hOLflat_ensemble}).
Assuming a flat universe, 
Stage IV time delays could provide independent evidence for $\OL > 0$.
Whether this remains interesting by Stage IV remains to be seen.
The constraints on $h$ are certainly tighter
and would be improved by the introduction of a prior on $\OL$,
which we defer until \S\ref{sec:cosmofish}.

\subsection{Curved universe with cosmological constant ($h, \Om, \OL, \Ok$)}
\label{sec:curvedcosmoconst}

\begin{figure}
\plotone{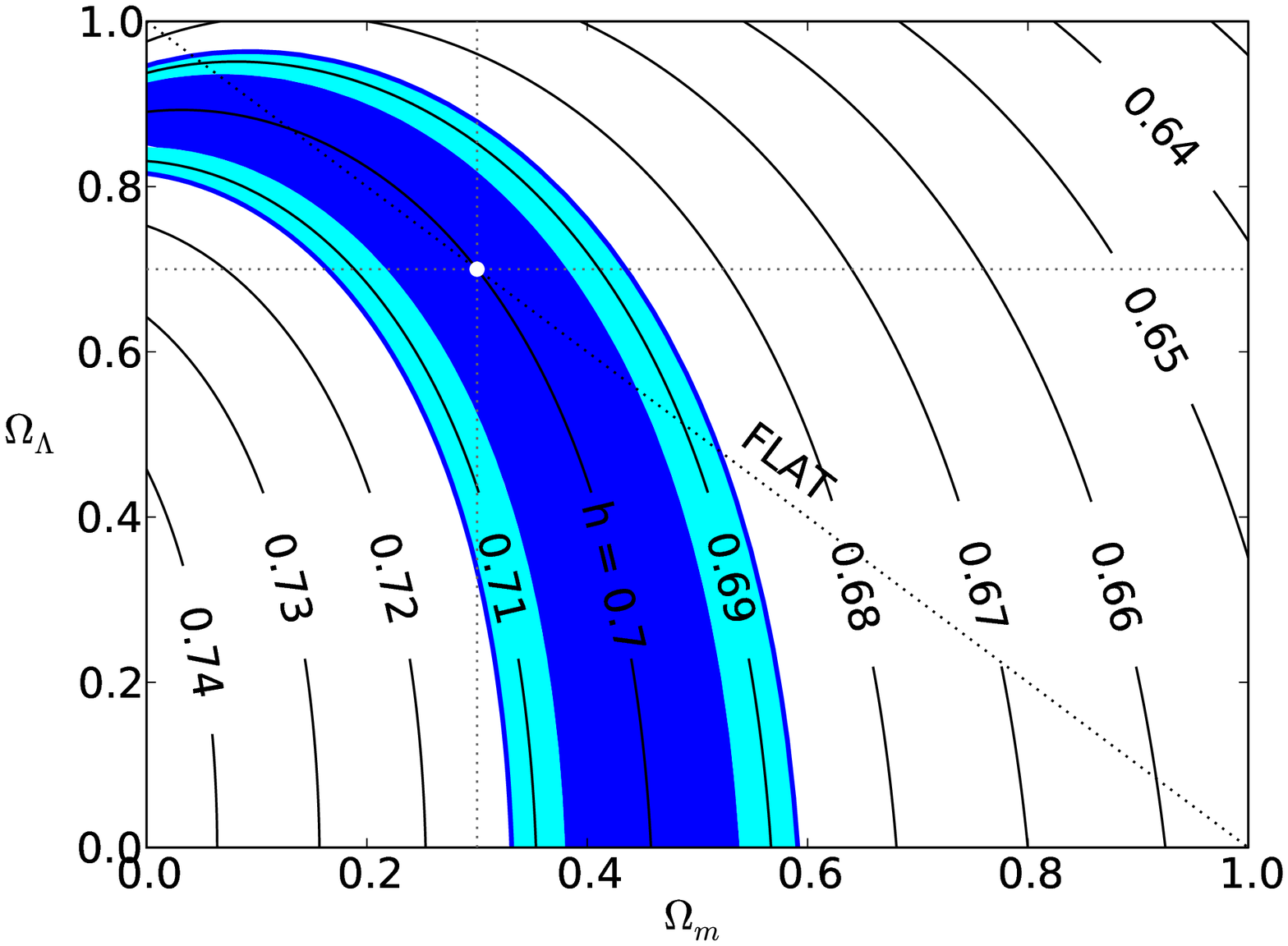}
\\
\caption[OmOLh]{
\label{fig:OmOLh}
Confidence contours (1- and 2-$\sigma$ colored bands) for ($\Om, \OL$)
given $\dTC = 0.64\%$ obtained from an ensemble with
all lenses and sources at $z_L$, $z_S$ = (0.5, 2.0).
The colored bands shift in ($\Om, \OL$) space as $h$ varies. 
A cosmological constant ($\w = -1$) is assumed.
The input cosmology ($h, \Om, \OL$) = (0.7, 0.3, 0.7)
is marked with a white dot.
Flat cosmologies lie along the dotted line,
and this line's intersection with the colored bands
explains the strange shape of the colored bands in the previous plot.
}
\end{figure}

If we relax the flatness parameter,
adding another free parameter $\Om$
(where curvature is determined by $\Ok = 1 - (\Om + \OL)$),
we run into the degeneracy in Fig.~\ref{fig:OmOLh}.
Plotted as colored bands
are the ($\Om$, $\OL$) confidence contours assuming constant $h = 0.7$
given $\delta \TC = 0.64\%$ from an ensemble 
with all lenses and sources at $z_L$, $z_S$ = (0.5, 2.0).
As $h$ varies, these contours move as shown.

\begin{figure}
\plotone{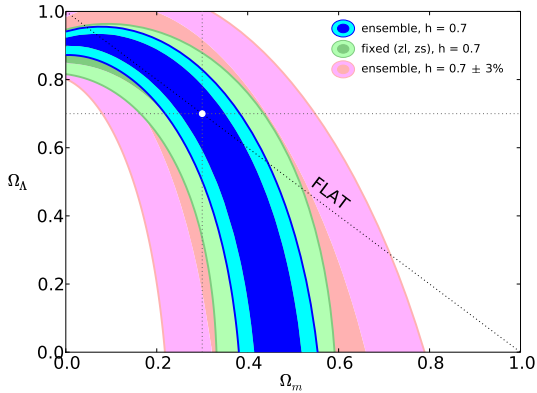}
\\
\caption[OmOLh]{
\label{fig:OmOLh_ensemble}
Additional confidence contours for ($\Om, \OL$).
The middle set of contours was plotted in the previous figure.
The top set of contours assumes an ensemble of lenses and sources
$z_L$, $z_S$ = ($0.5 \pm 0.15$, $2.0 \pm 0.75$).
Finally, the bottom set of contours is for the ensemble
and allowing a 3\% uncertainty in $h$.
}
\end{figure}

An ensemble of lenses with a range of redshifts
shrinks the confidence contours somewhat,
as we see in Fig.~\ref{fig:OmOLh_ensemble},
though the strong ($h, \Om, \OL$) degeneracy remains. 
Even adopting an aggressive 3\% prior on $h$, 
we find neither $\Om$ nor $\OL$ can be constrained individually.
However, the degeneracy does exhibit
a strong preference toward a flat or nearly flat universe.
Finally, we note the ($h, \Om, \OL$) degeneracy can be more cleanly broken
if our ensemble includes a significant fraction of lenses
at $z_L = 1$ and higher.

\subsection{Flat universe with constant dark energy EOS ($h, \Ode = 1 - \Om, \w$)}
\label{sec:flatwconst}

Current cosmological constraints are consistent with 
a flat universe with a cosmological constant
(as explored in \S\ref{sec:flatcosmoconst}).
As a first perturbation to this model,
it is common to explore constraints on $\w \neq -1$
while maintaining constant $\w$ in a flat universe.
This cosmology has three free parameters ($h, \Ode, \w$)
with $\Om = 1 - \Ode$.

\begin{figure}
\plotone{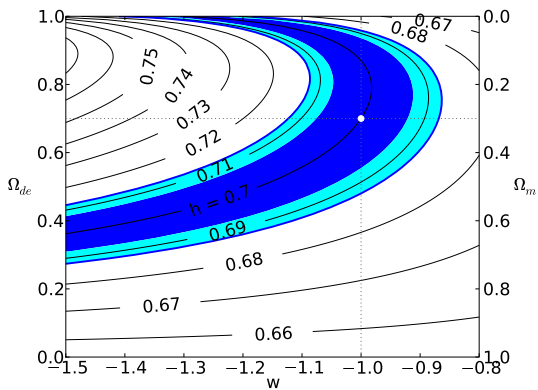}
\\
\caption[Hwflat]{
\label{fig:Hwflat}
Confidence contours (1- and 2-$\sigma$ colored bands) for ($\w, \Ode = 1 - \Om$)
assuming a flat universe with constant $\w$
given $\dTC = 0.64\%$ obtained from an ensemble with
all lenses and sources at $z_L$, $z_S$ = (0.5, 2.0).
The colored bands shift in ($\w, \Ode$) space as $h$ varies. 
The input cosmology ($h, \Ode, \w$) = (0.7, 0.7, -1)
is marked with a white dot.
}
\end{figure}

\begin{figure}
\plotone{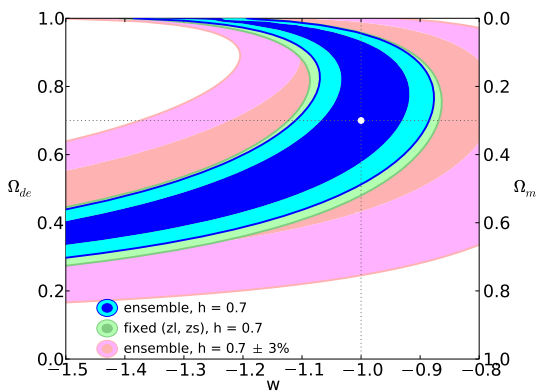}
\caption[Hwflat_ensemble]{
\label{fig:Hwflat_ensemble}
Confidence contours for ($\w$, $\Ode = 1-\Om$),
assuming a flat universe.
As in Fig.~\ref{fig:OmOLh_ensemble},
we plot a ``Stage IV'' ensemble of lenses at a range of redshifts,
the lenses all at the same redshift,
and the ensemble allowing 3\% uncertainty in $h$.
}
\end{figure}

Given enough data and appropriate priors,
time delay lenses could place strong constraints on
the dark energy equation of state parameter $\w$ (see \S\ref{sec:Fisherflatwconst}).
Figs.~\ref{fig:Hwflat} and \ref{fig:Hwflat_ensemble} 
explore the dependence of
$\TC$ on $(\w, \Ode)$ assuming a flat universe and constant $\w$.

\subsection{Flat universe with time-variable dark energy EOS 
($h, \Ode = 1 - \Om, \wo, \wa$)}
\label{sec:flatwvar}

\begin{figure*}
\plottwo{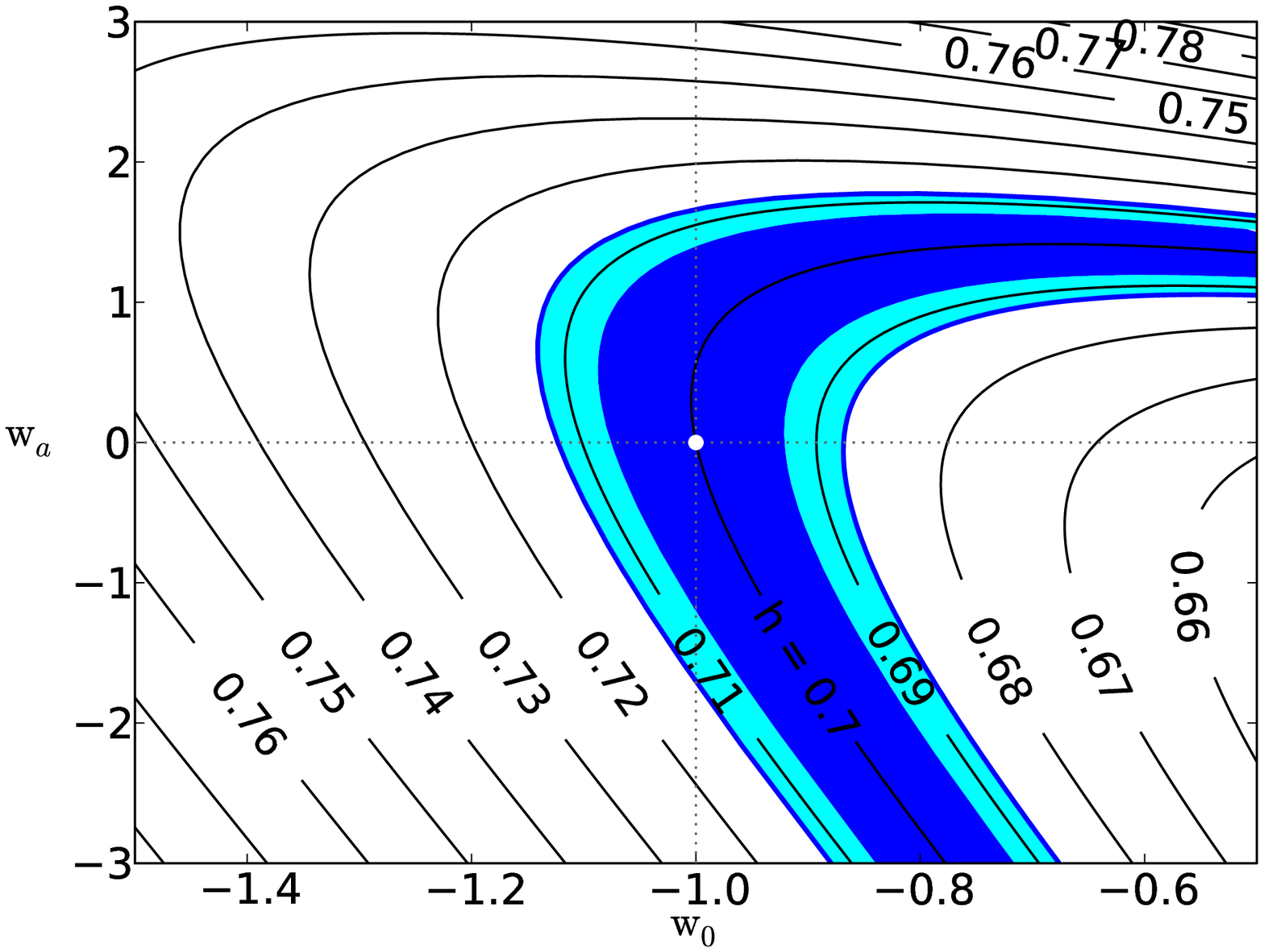}{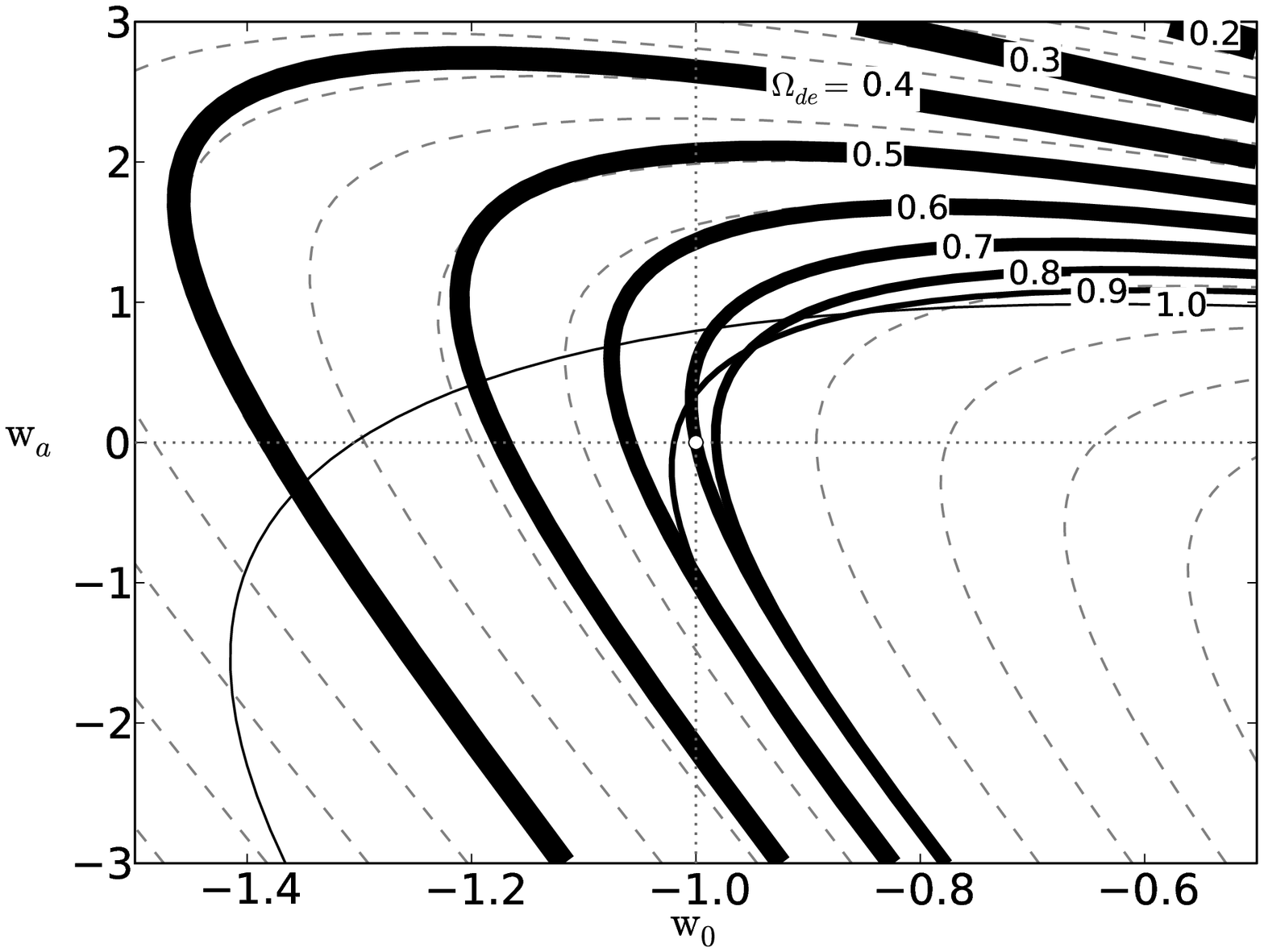}
\caption[w0wah]{
\label{fig:w0wah}
\textit{Left}: Confidence contours (1- and 2-$\sigma$ colored bands) for ($\wo, \wa$)
given $\dTC = 0.64\%$ obtained from an ensemble with
all lenses and sources at $z_L$, $z_S$ = (0.5, 2.0)
and assuming $h = 0.7$ and perfect knowledge of ($\Om, \Ode$).
As shown, these bands shift in ($\wo, \wa$) space as $h$ varies.
The input cosmology ($h, \wo, \wa$) = (0.7, -1, 0)
is marked with a white dot.
\textit{Right:} Dependence of the ($\wo, \wa$) contours on $\Ode$,
assuming a flat cosmology.
Dashed lines show the $h$ dependence from the left plot.
Solid lines of increasing thickness show contours of $\Ode$ 
decreasing in 0.1 increments.
}
\end{figure*}

The most interesting constraints we can hope to place on dark energy are
to verify or falsify the following:
$\w = -1$ (cosmological constant) and
$\wa = 0$ (constant $\w$).
In Fig.~\ref{fig:w0wah} we explore the dependence of 
$\TC$ on $(\wo, \wa)$ (see Eq.~\ref{eq:w}).
The colored bands are the constraints we could obtain
given perfect knowledge of ($h, \Om, \Ode$).
The solid lines on the left show the curves' migration
as a function of $h$.
On the right, we also explore dependence on $\Ode$
for a flat universe ($\Om + \Ode = 1$).

\section{Cosmological Constraints from Future Experiments}
\label{sec:cosmofish}

We now consider the full parameter space $(h, \Om, \Ode, \Ok, \wo, \wa)$
and derive the constraints that may be placed on these parameters
given constraints on $\TC$ along with various priors.
Stage IV time delay constraints are compared to those expected from other experiments
as estimated by the Dark Energy Task Force \citep{DETF06,DETF09}.
To efficiently explore this parameter space, we perform Fisher matrix analyses.

\subsection{Fisher Matrix Analysis}
\label{sec:Fisher}

The Fisher matrix formalism
provides a simple way to study uncertainties of many correlated parameters.
Constraints from various experiments and/or specific priors
may be combined with ease.
A ``quick-start'' instructional guide and software 
are provided in a companion paper \citep{Coe09Fisher}.
Fisher matrices approximate all uncertainties as Gaussians.
The true uncertainties may be somewhat higher and non-Gaussian.
The full information of the dependencies as shown in \S\ref{sec:cosmo} is not retained.
Yet as cosmological parameters are constrained close to their true values,
these approximations should suffice.

As above we consider a ``Stage IV'' ensemble of time delays
which constrains $\TC$ to 0.64\%
with Gaussian distributions of lens and source redshifts
($z_L = 0.5 \pm 0.15$;
$z_S = 2.0 \pm 0.75$).
Assuming such a Gaussian distribution for $\TC$ and the aforementioned redshift ensemble,
we calculate (numerically) the Fisher matrix for cosmological parameters of interest.
The Fisher matrix consists of partial derivatives of $\chi^2$ 
with respect to the parameters.
For parameters ($p_i, p_j$), element ($i, j$) in the Fisher matrix is given by
\\
\begin{equation} 
  F_{ij} = \frac{1}{2} \frac{\p \chi^2}{\p p_i \p p_j}.
\end{equation}

The Stage IV ($\dTC = 0.64\%$) 
time delay Fisher matrix is given in Table \ref{tab:fish}
for the cosmological parameters
$(h, \Ode, \Ok, \wo, \wa)$.
The Fisher matrix may be easily scaled to other $\dTC$ values.
For example, to scale from 
LSST (4,000 lenses; $\dTC = 0.64\%$) to
Pan-STARRS 1 (1,000 lenses; $\dTC = 1.27\%$),
simply divide all the values in the Fisher matrix by 
$4 = 4,000 / 1,000 = (1.27 / 0.64)^2$.
Or multiply them by $1.6 = (0.64 / 0.4)^2$
to explore the LSST + OMEGA constraints ($\dTC = 0.4\%$).
If one is interested in constraints on 
$\Om = 1 - (\Ode + \Ok)$,
$\om \equiv \Om h^2$,
or any other related variable,
a transformation of variables can be performed as outlined in \cite{Coe09Fisher}.

\begin{deluxetable}{lrrrrr}
\tablecaption{Stage IV Fisher matrix expectation for $(h, \Ode, \Ok, \wo, \wa)$\\
given $\dTC = 0.64\%$ \label{tab:fish}}
\tablewidth{0pt}
\tablehead{
& \colhead{$\h$}
& \colhead{$\Ode$}
& \colhead{$\Ok$}
& \colhead{$\wo$}
& \colhead{$\wa$}
} 
\startdata
$\h$ & 49824.9224 & -1829.7018 & -4434.2995 & 4546.8899 & 122.5319 \\
$\Ode$ & -1829.7018 & 88.3760 & 200.9795 & -189.2658 & -8.4386 \\
$\Ok$ & -4434.2995 & 200.9795 & 463.5732 & -445.5690 & -17.9694 \\
$\wo$ & 4546.8899 & -189.2658 & -445.5690 & 441.9725 & 15.2981 \\
$\wa$ & 122.5319 & -8.4386 & -17.9694 & 15.2981 & 1.0394 \\
\vspace{-0.1in}
\enddata
\end{deluxetable}
 
\begin{figure*}
\plotone{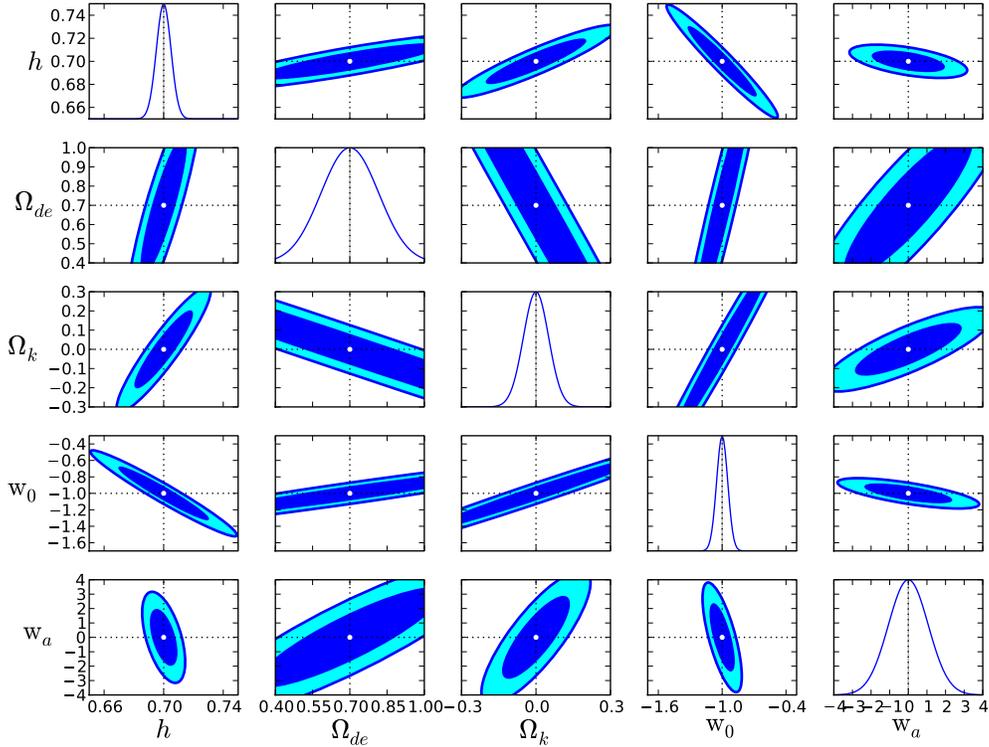}
\\
\caption[hcosmo]{
\label{fig:Fish}
Constraints placed on pairs of parameters
derived from our Fisher matrix analysis
assuming perfect knowledge of all other parameters
given $\dTC = 0.64\%$ obtained from an ensemble with
Gaussian distributions of lens and source redshifts
($z_L = 0.5 \pm 0.15$;
$z_S = 2.0 \pm 0.75$).
All off-diagonal plots show 1- and 2-$\sigma$ colored ellipses.
Along the diagonal are constraints on individual parameters
assuming perfect knowledge of all others.
The y axes along the diagonal are units of relative probability,
different from the off-diagonal plots.
}
\end{figure*}

In Fig.~\ref{fig:Fish} we show the time delay constraints possible 
on all parameters and pairs of parameters
assuming perfect knowledge of all the other parameters.
These plots can be compared to those presented in \S\ref{sec:cosmo}.
Such perfect priors are unrealistic,
but they help to demonstrate the parameter dependencies and degeneracies.

\subsection{Flat universe with constant $\w$}
\label{sec:Fisherflatwconst}

We first consider the simple case of a flat universe with constant $\w$.
This is a common perturbation to the concordance cosmology.
The goal is to detect deviation from $\w = -1$, 
equivalent to the cosmological constant $\Lambda$.
This 3-parameter cosmology ($\h, \Ode, \w$, with $\Om = 1 - \Ode$)
was explored above in \S\ref{sec:flatwconst}.

\begin{figure*}
\includegraphics[width=0.33\hsize]{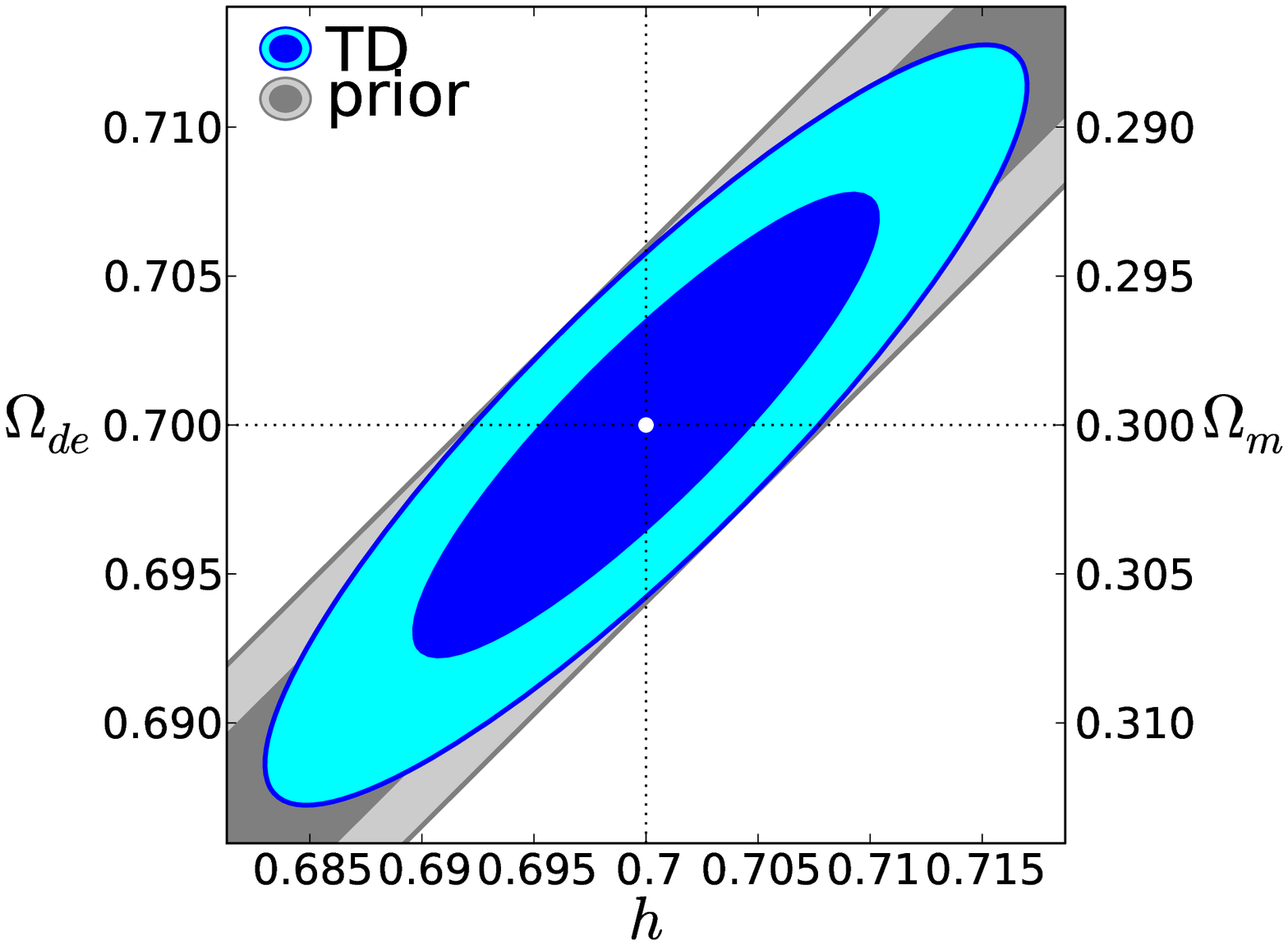}
\includegraphics[width=0.33\hsize]{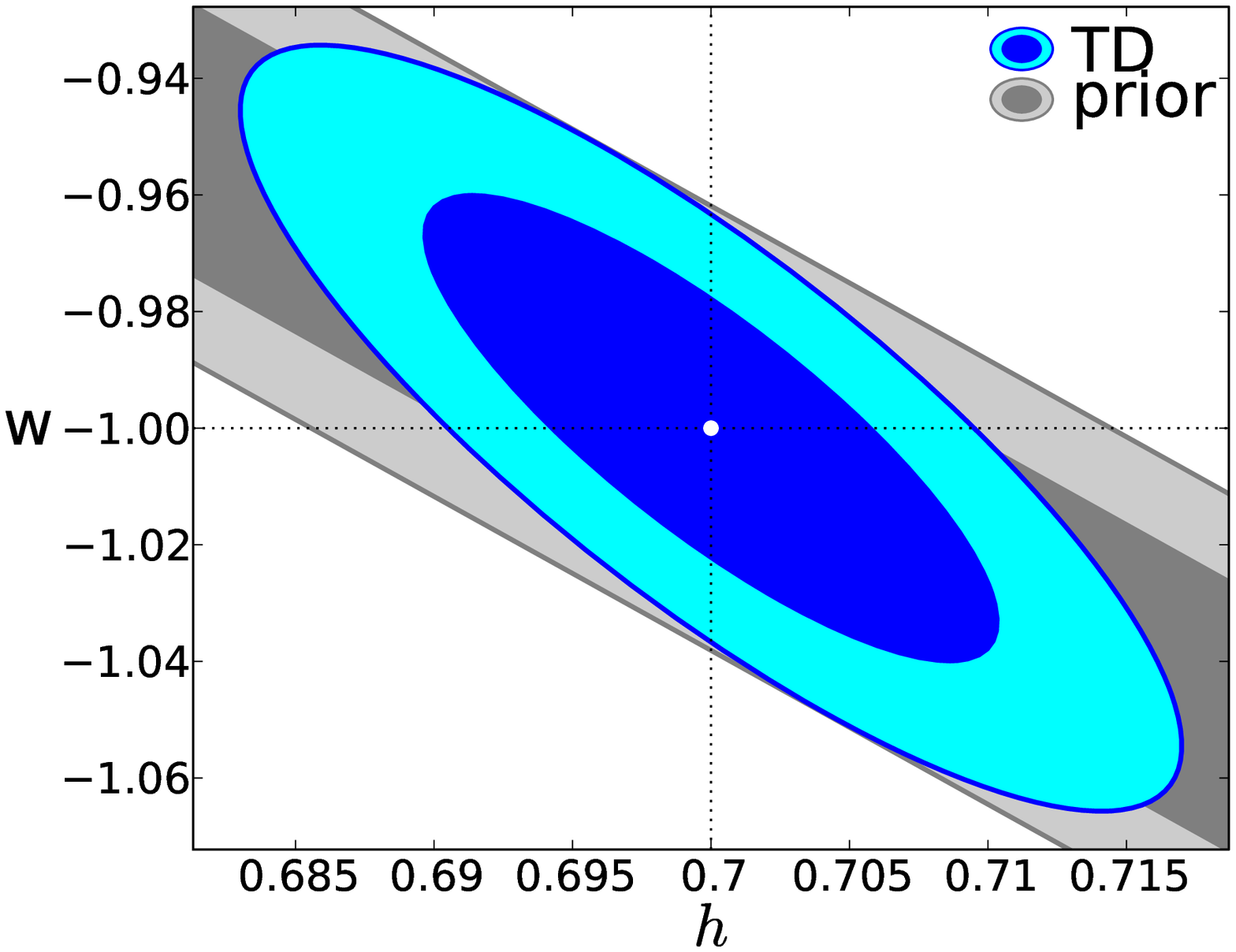}
\includegraphics[width=0.33\hsize]{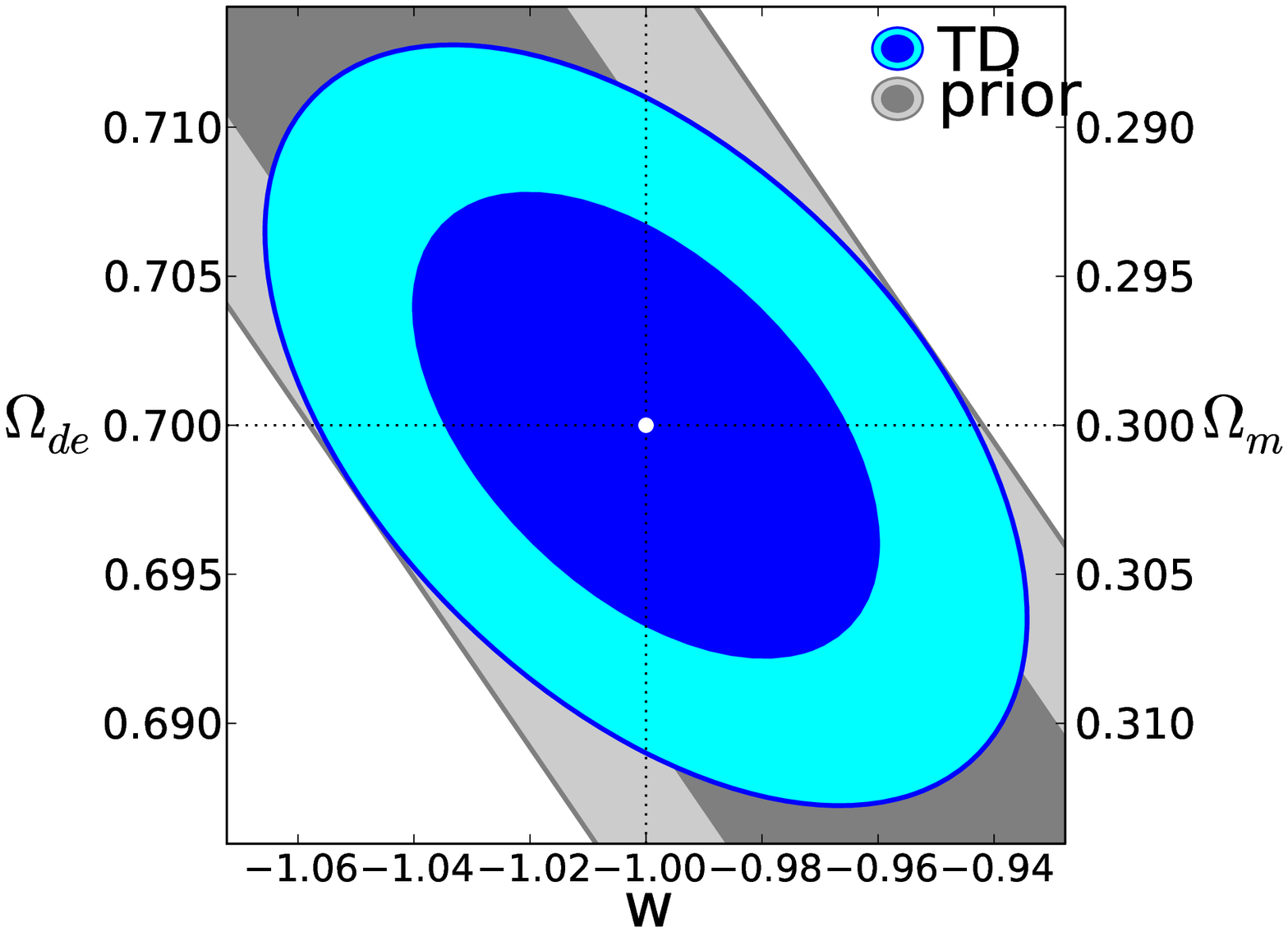}
\includegraphics[width=0.33\hsize]{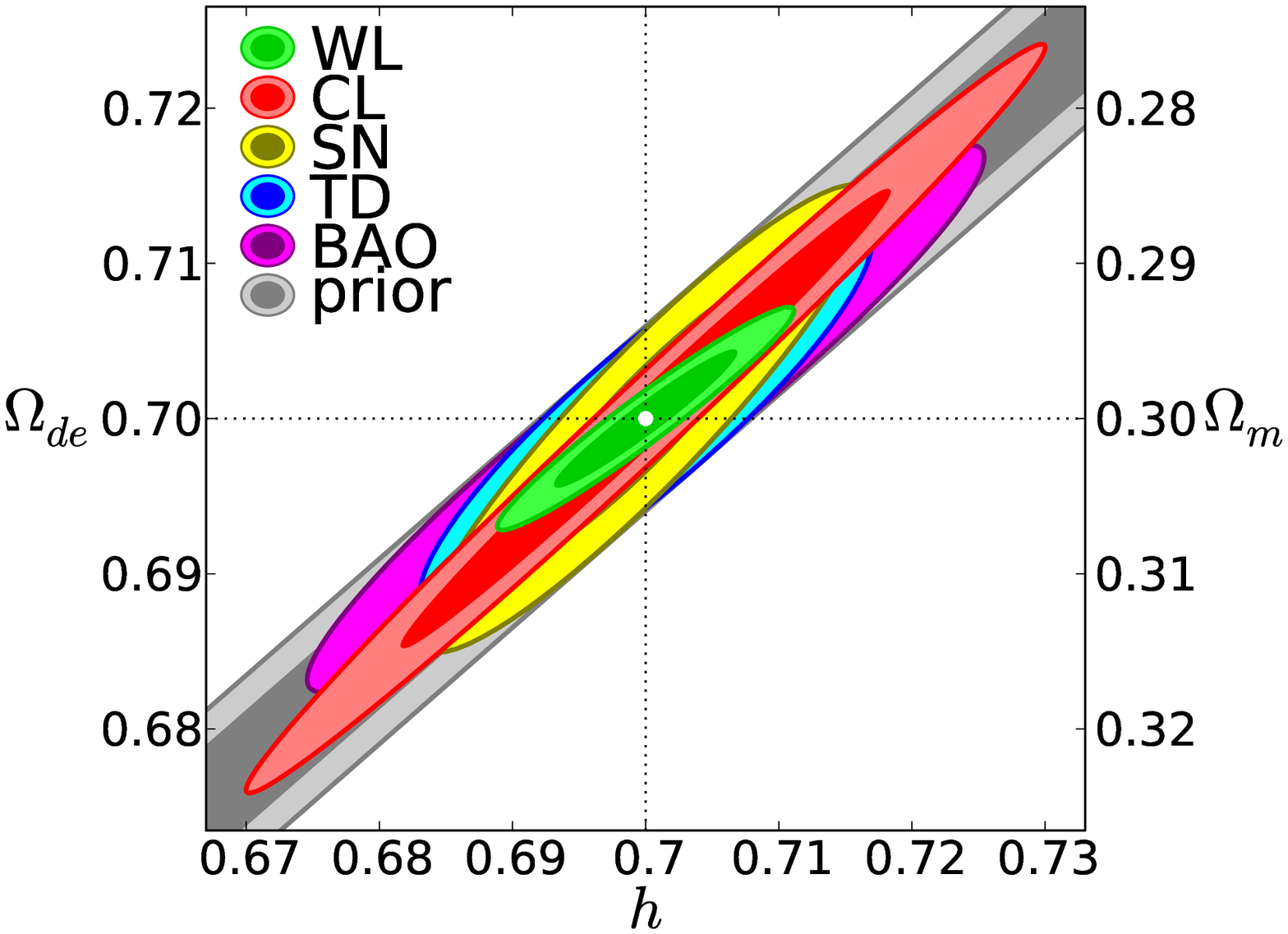}
\includegraphics[width=0.33\hsize]{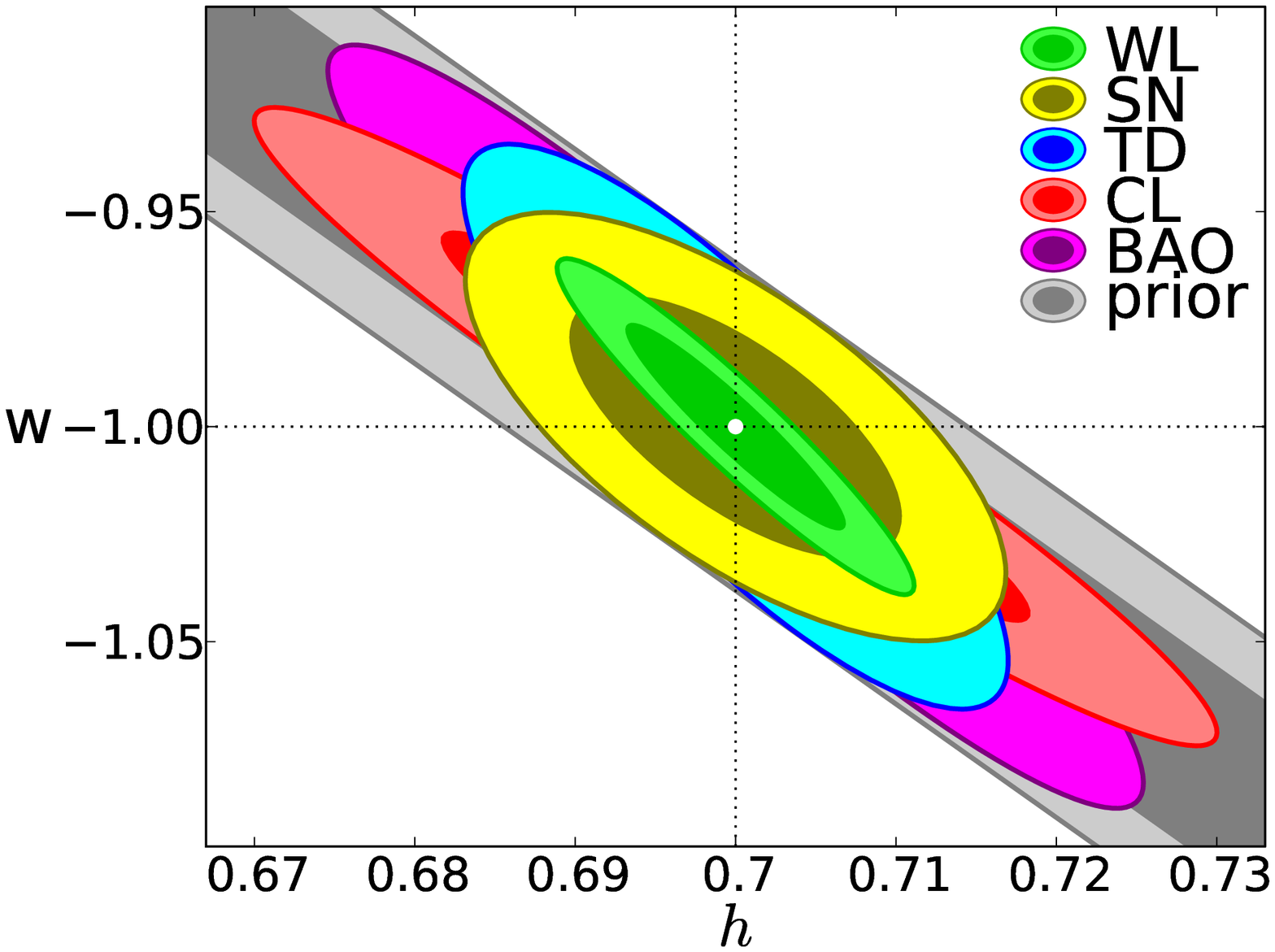}
\includegraphics[width=0.33\hsize]{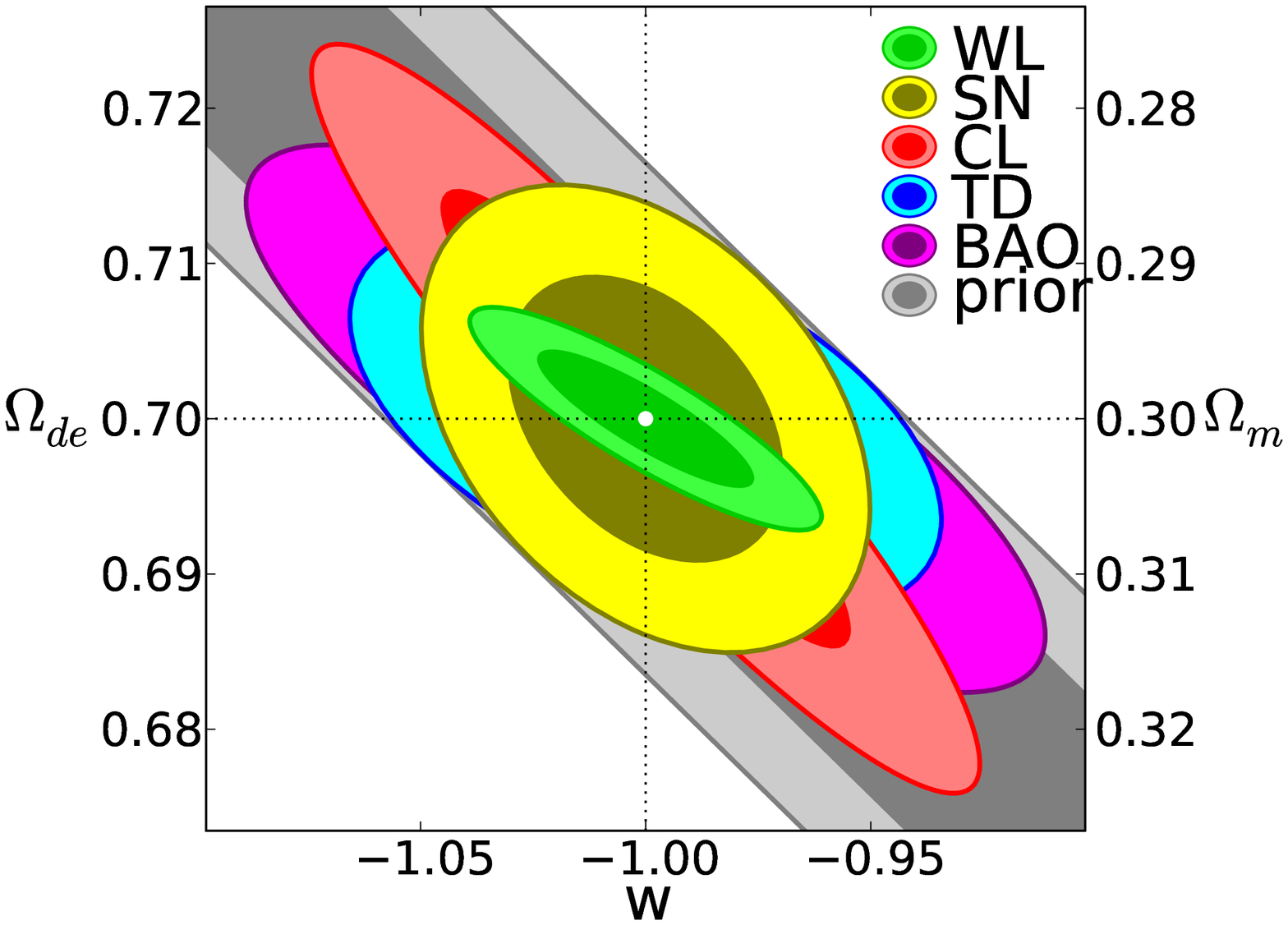}
\caption[StageIVflatwconst]{
\label{fig:StageIVflatwconst}%
{\it Top row:} Cosmological constraints from 
``Stage IV'' time delays plus a Planck prior in a flat universe with constant $\w$.
We assume an ensemble of time delays which constrains $\TC$ to 0.64\% 
(see text for details).
Time delays plus Planck constrain 
$h$ to 0.007 (1\%), $\Ode$ to 0.005, and $\w$ to 0.026 
(all 1-$\sigma$ precisions).
{\it Bottom row:} Comparison of ``optimistic Stage IV'' constraints expected from 
time delays (TD), 
weak lensing (WL), 
supernovae (SN), 
baryon acoustic oscillations (BAO),
and cluster counts (CL).
The time delay constraints are as plotted in the top row.
For the other experiments
we use Fisher matrix calculations provided by the Dark Energy Task Force (DETF).
For each parameter pair,
experiments are plotted in order of ${\rm FOM} \propto ({\rm Ellipse ~ Area})^{-1}$,
with the best experiment on top.
}
\end{figure*}

The top row of Fig.~\ref{fig:StageIVflatwconst}
shows Stage IV time delay constraints with a Planck prior
in a flat universe with constant $\w$.
Given these priors, we estimate that time delays will constrain
$h$ to 0.007 ($\sim 1\%$), $\Ode$ to 0.005, and $\w$ to 0.026 (all 1-$\sigma$ precisions).

In the bottom row of Fig.~\ref{fig:StageIVflatwconst},
we compare these time delay constraints (TD)
to those expected from other methods: 
weak lensing (WL),
baryon acoustic oscillations (BAO),
supernovae (SN),
and
cluster counts (CL).
We consider ``optimistic Stage IV'' expectations from these methods
as calculated by the Dark Energy Task Force (DETF; \citealt{DETF06,DETF09})
and made available in the software
DETFast\footnote{\tt http://www.physics.ucdavis.edu/DETFast/}.
A Planck prior (also calculated by the DETF) is again assumed for all experiments.

In manipulating the DETF Fisher matrices we adopt their cosmology
($\Om, \Ode, h$) = (0.27, 0.73, 0.72),
but we revert to our chosen cosmology
($\Om, \Ode, h$) = (0.3, 0.7, 0.7)
for the rest of our analysis.
These differences have negligible impact on our results.

\cite{LewisIbata02} considered similar constraints from time delay lenses
but found much weaker constraints on ($\h, \w$),
even with all other cosmological parameters fixed.
One of the cases they considered was 500 lenses with 15\% uncertainty each,
which translates to $15\% / \sqrt{500} = 0.66\%$ total uncertainty,
very similar to the 0.64\% uncertainty we estimate for LSST
given 4,000 lenses with a much higher uncertainty (effectively 40\%) assumed per lens.
For this case, they find $0.99 \lesssim h \lesssim 1.10$ 
and $-1.48 \lesssim \w \lesssim -0.88$ (95\% confidence).
When we perform a similar analysis,
assuming $\dTC = 0.64\%$ and perfect knowledge of ($\Om, \Ode, \Ok, \wa$),
we obtain similar uncertainties (without biases, by construction): 
$\h = 0.7 \pm 0.02$ and $\w = -1 \pm 0.21$ (1-$\sigma$).
But with the addition of a Planck prior, 
even while relaxing the perfect prior on ($\Om, \Ode, \Ok, \wa$),
we find improved constraints of 
$\h = 0.7 \pm 0.007$ and $\w = -1 \pm 0.026$ (1-$\sigma$).
Planck clearly complements the strong lensing constraints well
to produce tight constraints on ($\h$, $\w$).

\subsection{General Cosmology}
\label{sec:cosmogen}

We now assume a general cosmology
allowing for curvature and a time-varying $\w$.
To help constrain this larger parameter space
($h, \Ode, \Ok, \wo, \wa$, with $\Om = 1 - (\Ode + \Ok)$),
we add additional priors.
In addition to the Planck prior,
we adopt ``Stage II'' (near-future) constraints
from weak lensing (WL) + supernovae (SN) + cluster counts (CL),
all as calculated by the DETF.
The DETF uses this prior (in addition to Planck)
in many of their calculations
comparing the performance of Stage III -- IV techniques.

The Stage II DETF WL + SN + CL prior yields the following uncertainties:
$\Delta h = 0.031$ (4.4\%),
$\Delta \Ode = 0.023$,
$\Delta \Ok = 0.010$,
$\Delta \wo = 0.128$,
$\Delta \wa = 0.767$
(along with various covariances between parameters).
The addition of the Planck prior reduces these to:
$\Delta h = 0.017$ (2.4\%),
$\Delta \Ode = 0.012$,
$\Delta \Ok = 0.003$,
$\Delta \wo = 0.115$,
$\Delta \wa = 0.525$.
Note that Stage II WL+SN+CL 
constrains $h$ well enough (to 4.4\%)
that an HST Key Project prior ($h = 0.72 \pm 0.08$) appears to be unnecessary.
Even SHOES ($h = 0.742 \pm 0.036$, or 4.9\%) 
provides a weaker constraint on $h$.
However, as noted in the introduction, these combined WL+SN+CL experiments
yield a prediction of $h$ based on an assumed cosmological model
and are no substitute for local measurements of $h$ \citep{Riess09}.

These Stage II constraints
are also rather optimistically combined,
assuming that all experiments have converged on the same best fit cosmology
without systematic offsets among them.
The true Stage II constraints should be somewhat weaker.

\begin{figure*}
\plotone{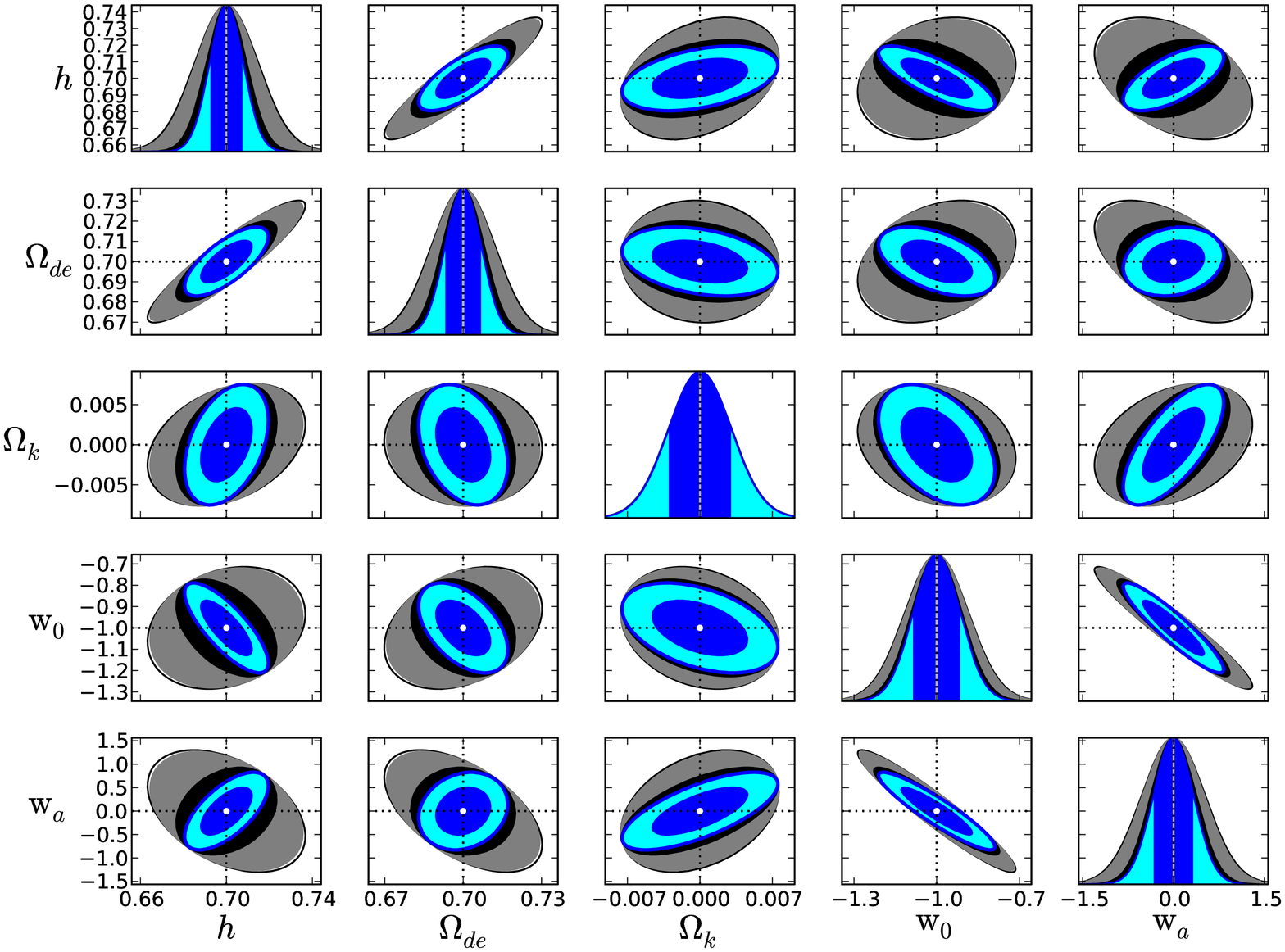}
\\
\caption[hcosmo]{
\label{fig:FishPlanck}
Cosmological constraints from time delays
in a general cosmology
assuming priors of Planck + Stage II (WL+SN+CL) 
as calculated by the DETF.
A progression is shown from the prior (outermost ellipse, 2-$\sigma$)
to Stage I (current) time delay constraints ($\dTC = 8.6\%$; gray ellipse, 2-$\sigma$)
to Stage II constraints ($\dTC = 1.4\%$; black ellipse, 2-$\sigma$)
to Stage IV constraints ($\dTC = 0.64\%$; colored ellipses, 1-$\sigma$ and 2-$\sigma$).
Along the diagonal are plotted constraints on individual parameters
marginalizing over all others.
The y axes along the diagonal are units of relative probability,
different from the off-diagonal plots.
For each plot, we marginalize over all other parameters,
unlike Fig.~\ref{fig:Fish} in which unrealistic perfect priors 
were assumed for illustrative purposes.
}
\end{figure*}

Plotted in Fig.~\ref{fig:FishPlanck} are
time delay constraints assuming a prior of Planck + Stage II WL+SN+CL.
A progression is shown
from Stage I (present) time delay constraints ($\dTC = 8.6\%$)
through Stage II ($\dTC = 1.27\%$)
and on to Stage IV ($\dTC = 0.64\%$).
The current constraints barely improve upon this aggressive prior.
While the Stage II -- IV constraints certainly improve upon the prior,
note that the outer bounds of the time delay and prior ellipses nearly intersect.
This indicates that the size of the time delay ellipse is controlled by that of the prior,
at least for these constraints and prior.
Were the prior significantly weaker 
or the time delay constraints significantly stronger,
we have verified that the time delay ellipses would shrink well within the prior ellipses.

\begin{figure*}
\includegraphics[width=0.45\hsize]{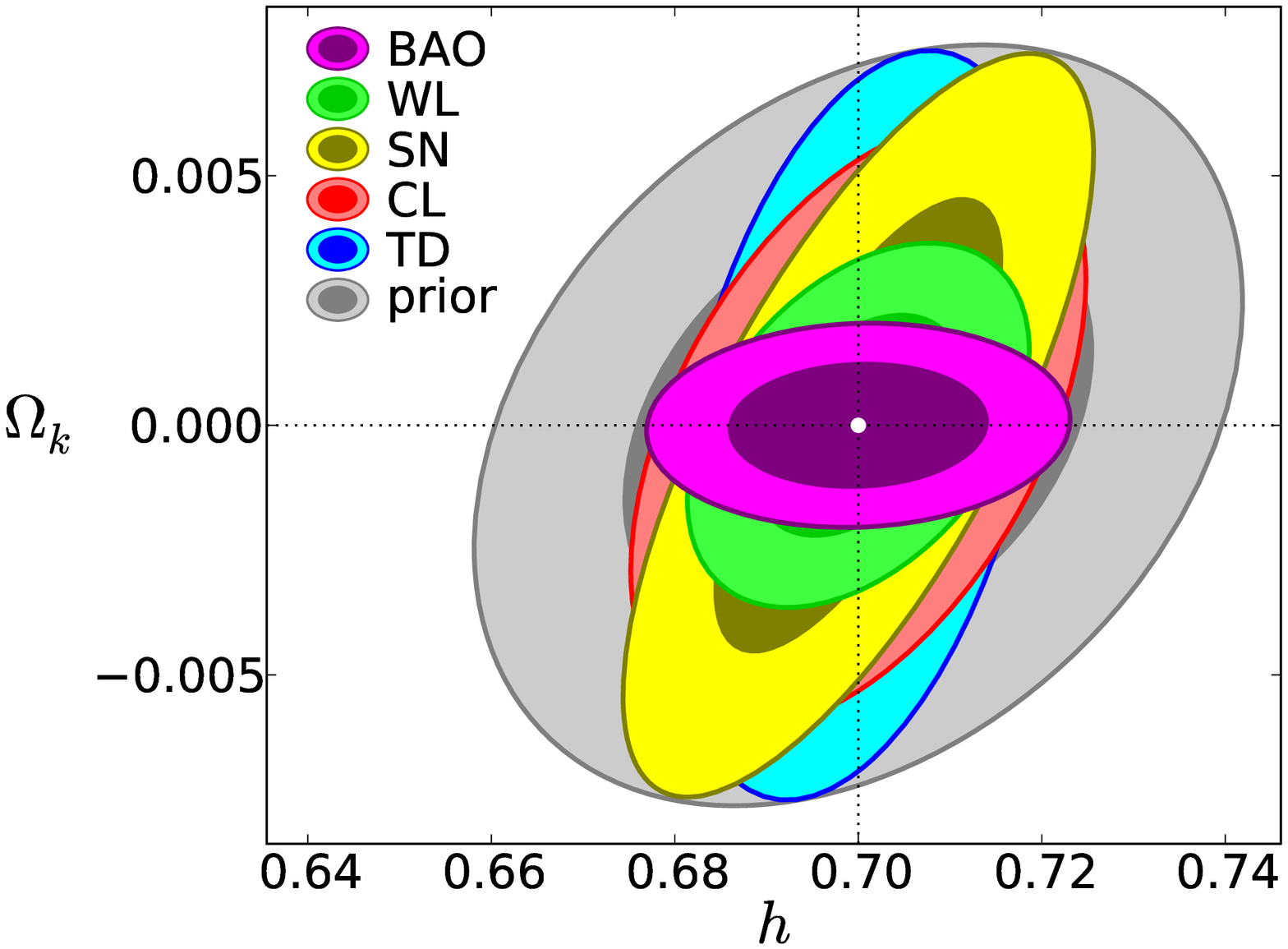}
\includegraphics[width=0.45\hsize]{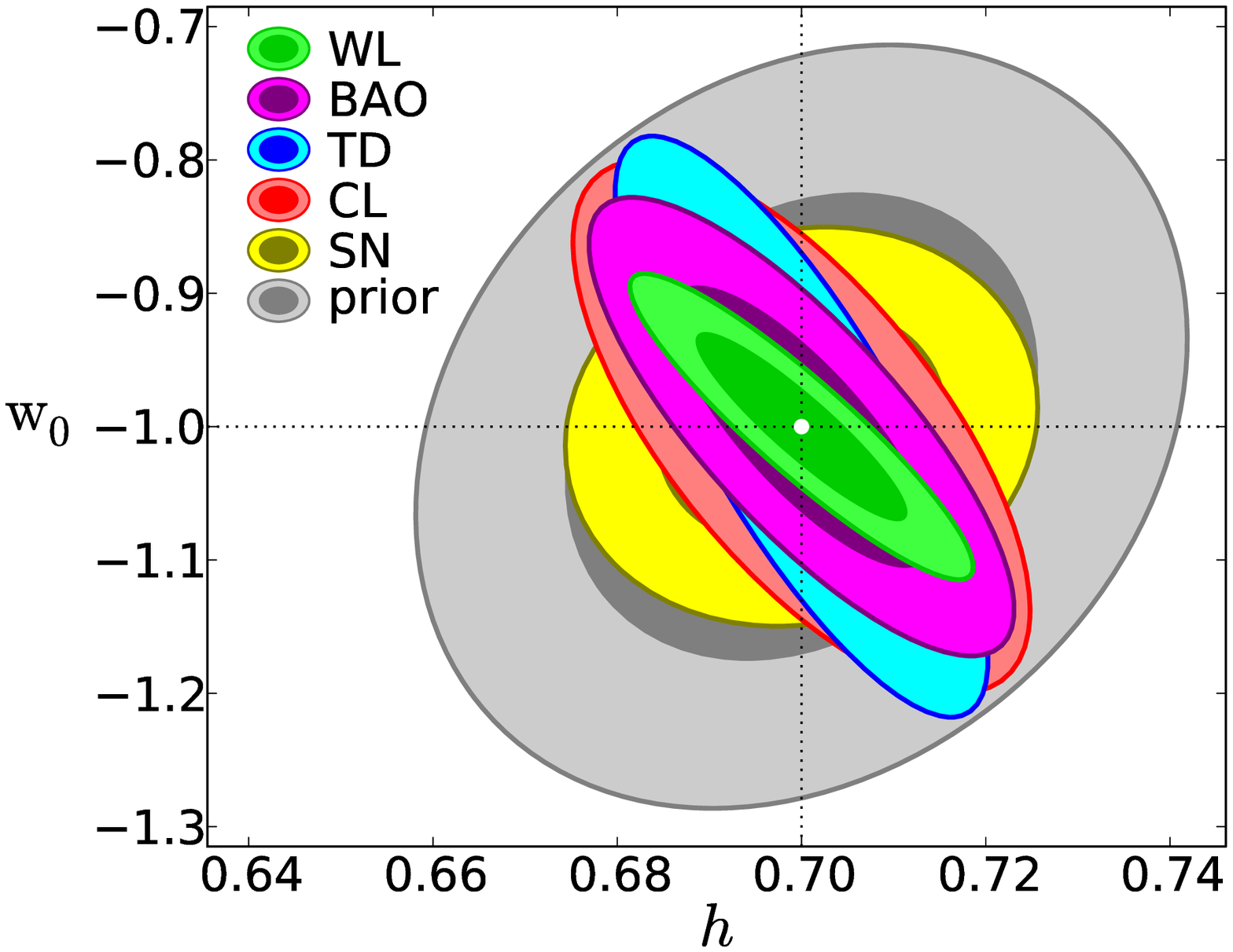}
\\
\includegraphics[width=0.45\hsize]{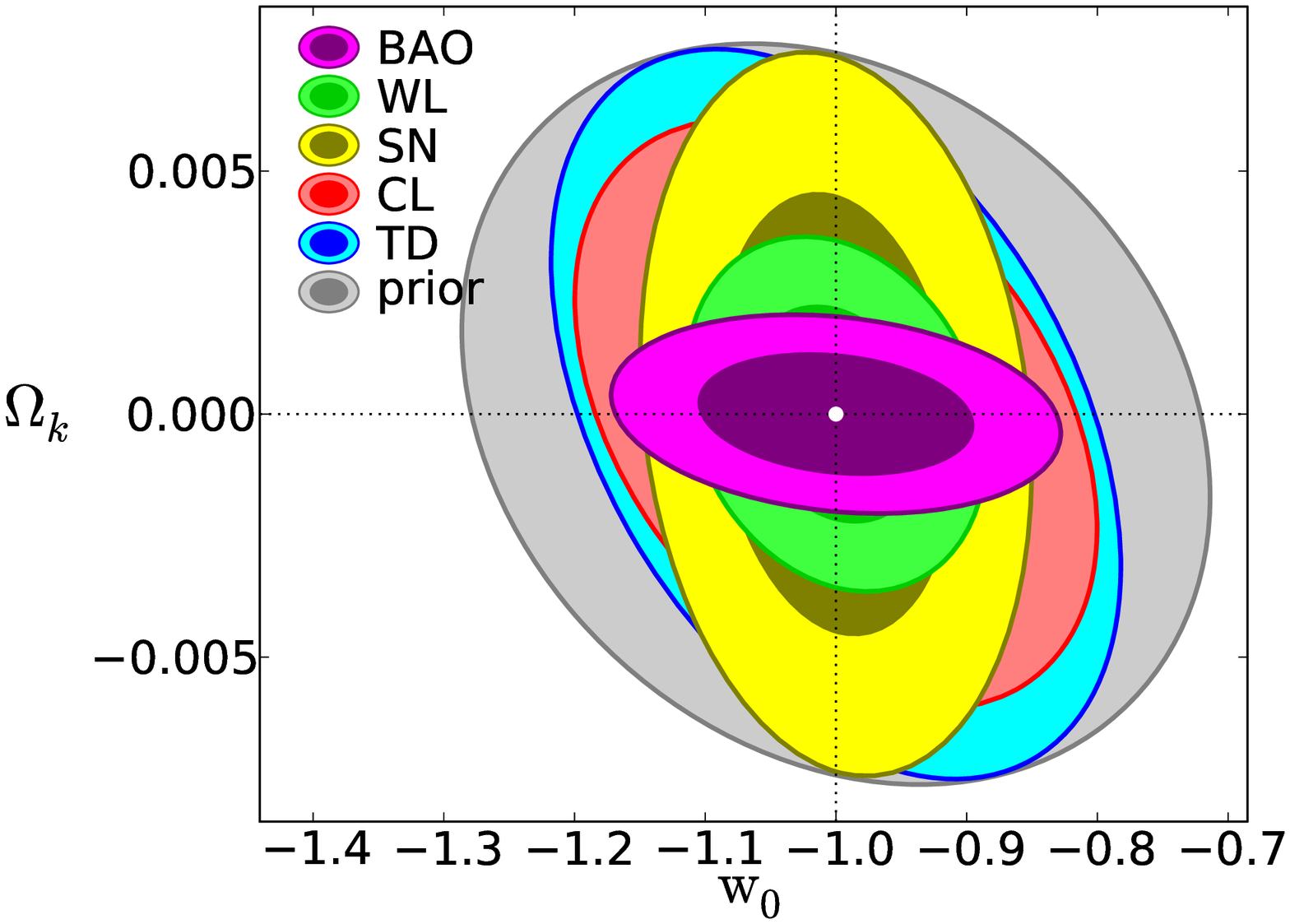}
\includegraphics[width=0.45\hsize]{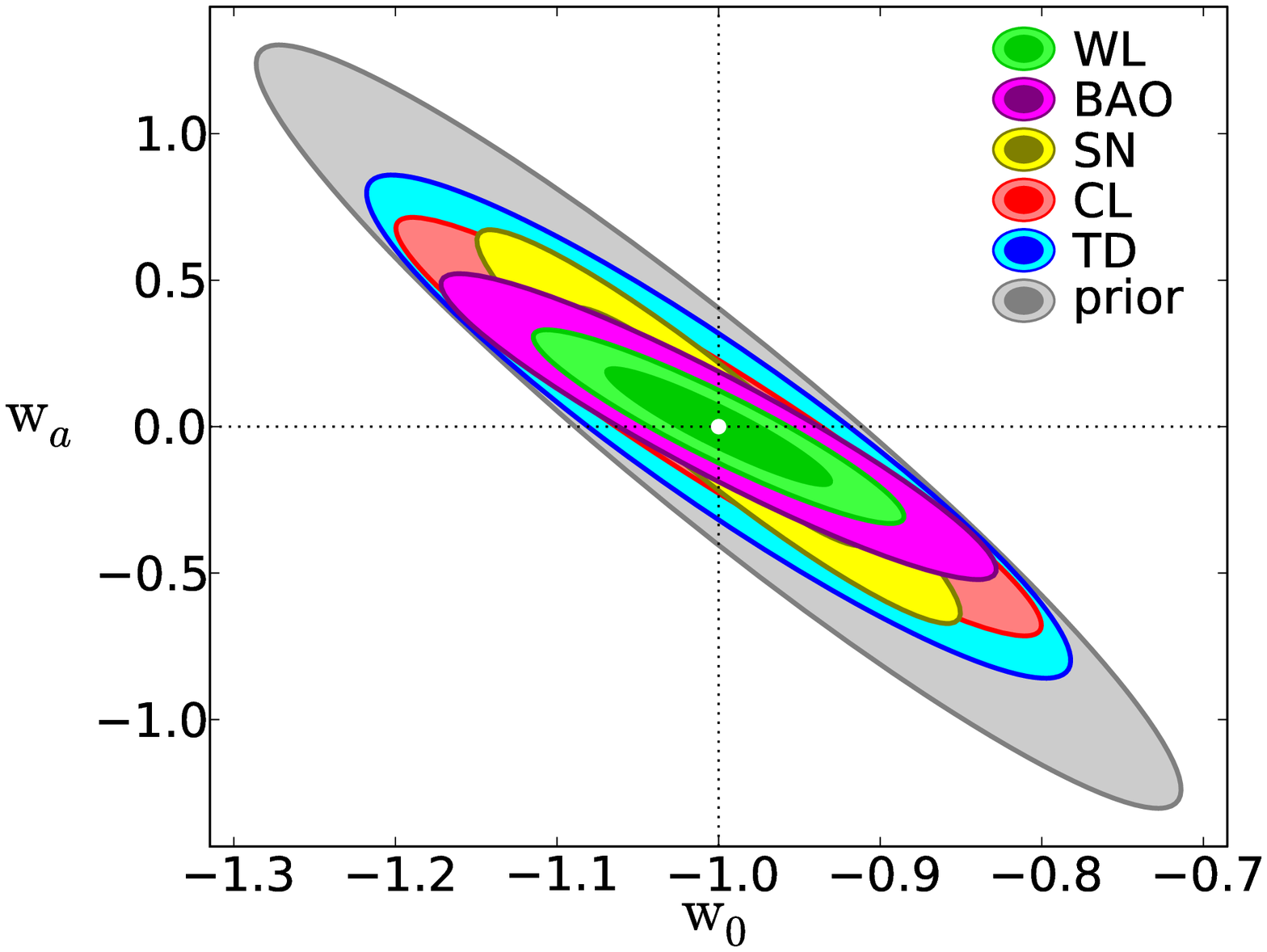}
\caption[StageIV]{
\label{fig:StageIV}%
Comparisons of ``Stage IV'' constraints possible from 
time delays (TD), 
weak lensing (WL), 
supernovae (SN), 
baryon acoustic oscillations (BAO),
and cluster counts (CL)
in a general cosmology (allowing for curvature and a time-variable $\w$).
For TD, we assume an ensemble which constrains $\TC$ to 0.64\% 
(see text for details).
For the rest we use ``optimistic Stage IV'' expectations calculated from 
Fisher matrices provided by the Dark Energy Task Force (DETF).
A prior of 
Planck + Stage II (WL+SN+CL) 
is assumed for all five experiments
and is plotted in gray.
For each parameter pair,
experiments are plotted in order of ${\rm FOM} \propto ({\rm Ellipse ~ Area})^{-1}$,
with the best experiment on top.
}
\end{figure*}

In Fig.~\ref{fig:StageIV} we compare Stage IV time delay constraints
to those expected from other methods
for various parameters of interest.
Plotted are constraints on ($\h, \Ok$), ($\h, \wo$), and ($\wo, \Ok$), and ($\wo, \wa$).
An example of how these constraints combine is given in \S\ref{sec:morethanh}.


\begin{figure}
\plotone{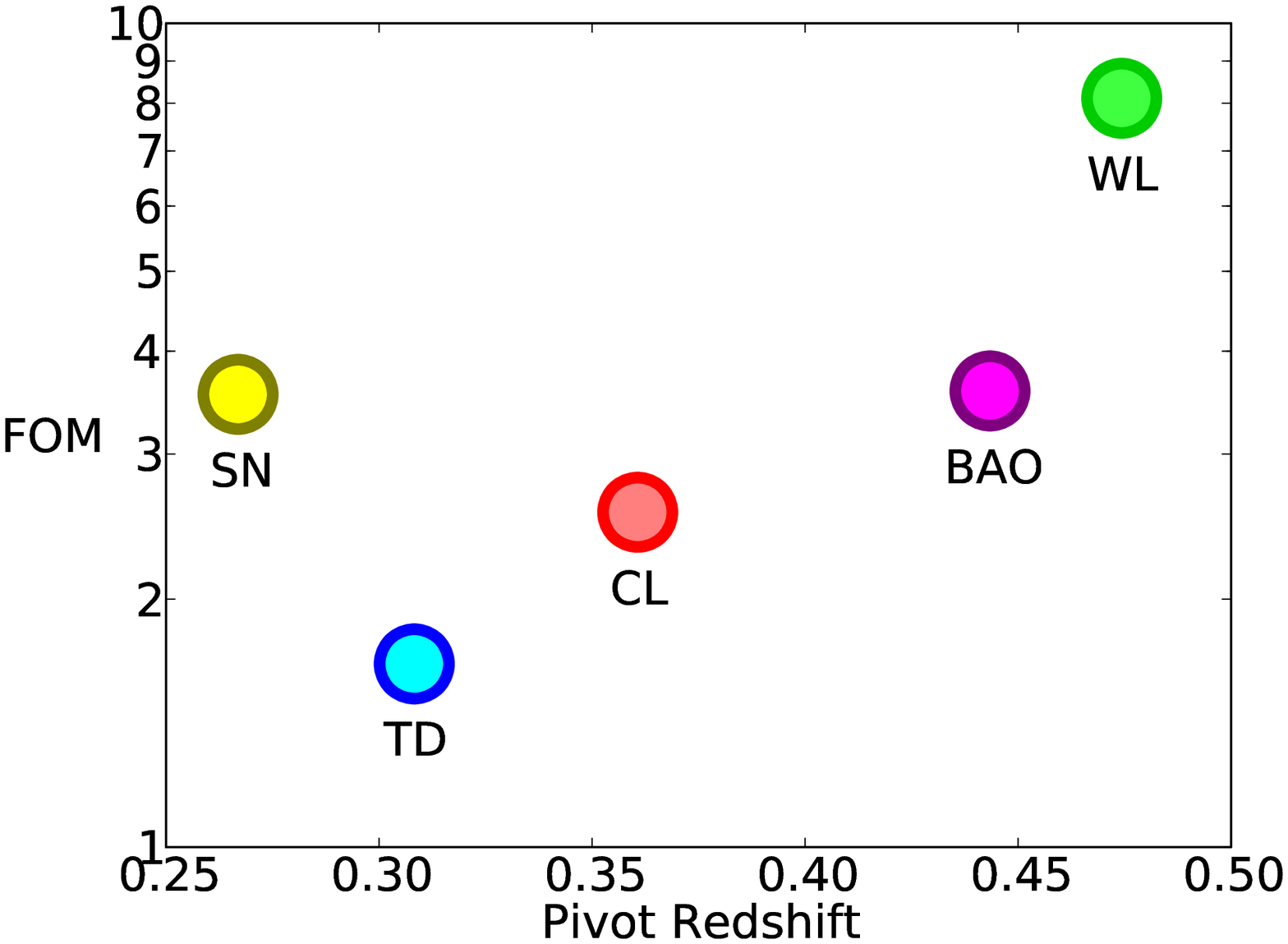}
\caption[FOMzp]{
\label{fig:FOMzp}%
Dark energy figure of merit 
(${\rm FOM} \propto \left((\wo,\wa) ~ {\rm  Ellipse ~ Area}\right)^{-1}$, 
normalized relative to the prior)
versus pivot redshift
for various ``optimistic Stage IV'' experiments 
with a prior of Planck + Stage II (WL+SN+CL)
The pivot redshift is the redshift at which $\w(z)$ is best constrained.
}
\end{figure}

We give extra attention to constraints on the dark energy parameters ($\wo, \wa$).
The DETF figure of merit (FOM) for a given experiment is defined as
the inverse of the area of the ellipse in the ($\wo, \wa$) plane.
In Fig.~\ref{fig:FOMzp} we plot FOM for various experiments 
versus the ``pivot redshift'', defined as follows.
For a time-varying $\w(z)$, time delays constrain $\w$ best at $z \approx 0.31$. 
This redshift is known as the pivot redshift \citep{HutererTurner01,HuJain04} 
and can also be calculated simply from the ($\wo, \wa$) constraints \citep{Coe09Fisher}. 
As in the previous plot, 
we assume a prior of Planck + Stage II (WL+SN+CL).

\subsection{Time delays do not simply constrain $h$}
\label{sec:hpriors}

\subsubsection{Relaxing the ``perfect prior'' on $(\Om, \Ode, \Ok, \wo, \wa)$}
\label{sec:relax}

To date, analyses of time delay lenses
have quoted uncertainties on $\TC$ as uncertainties on $h$,
assuming $\dh = \dTC$.
This assumption has been valid to date,
but future constraints on $h$ 
will be weaker than the constraints on $\TC$,
that is $\dh > \dTC$.

\begin{figure*}
\plottwo{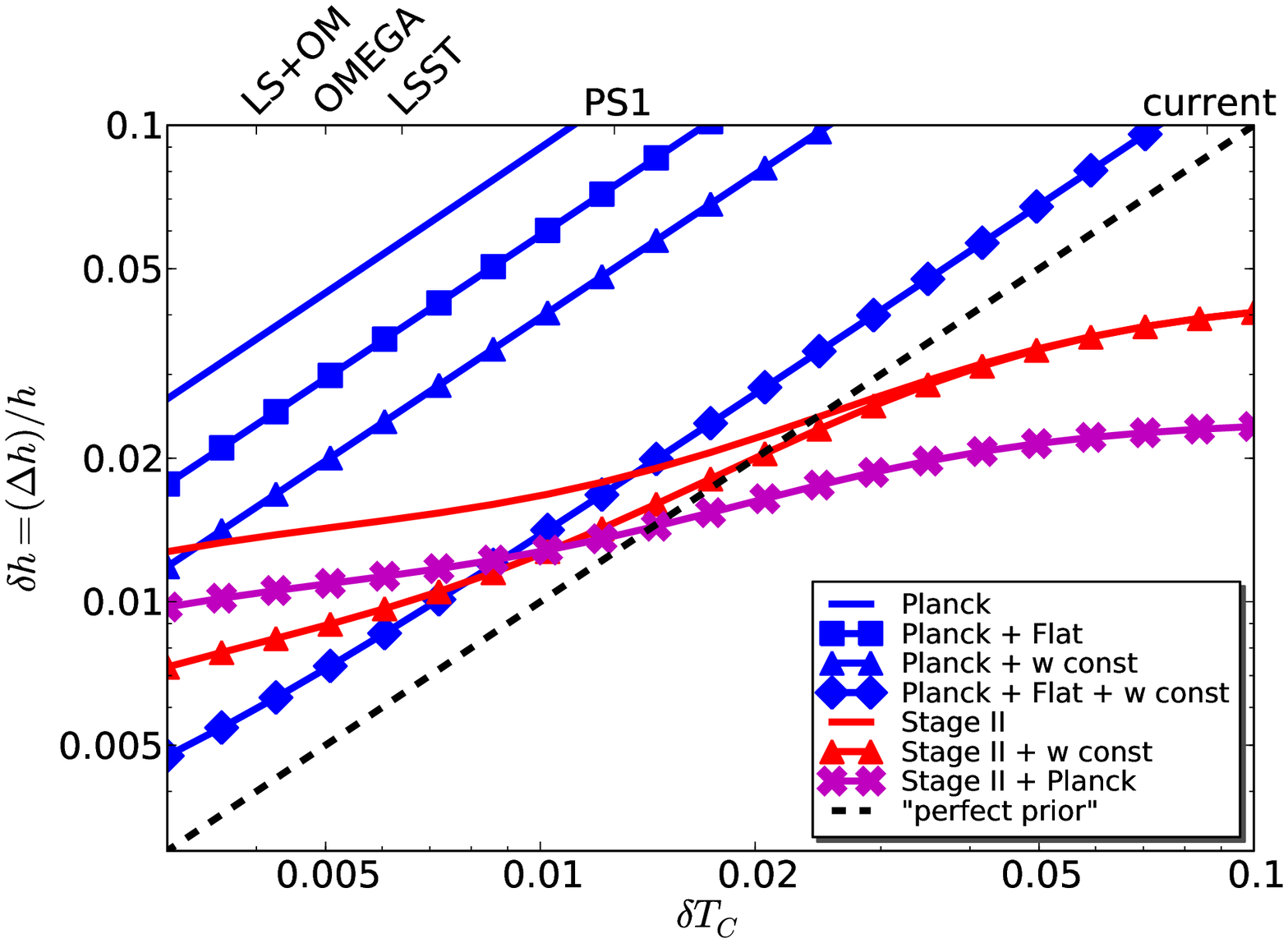}{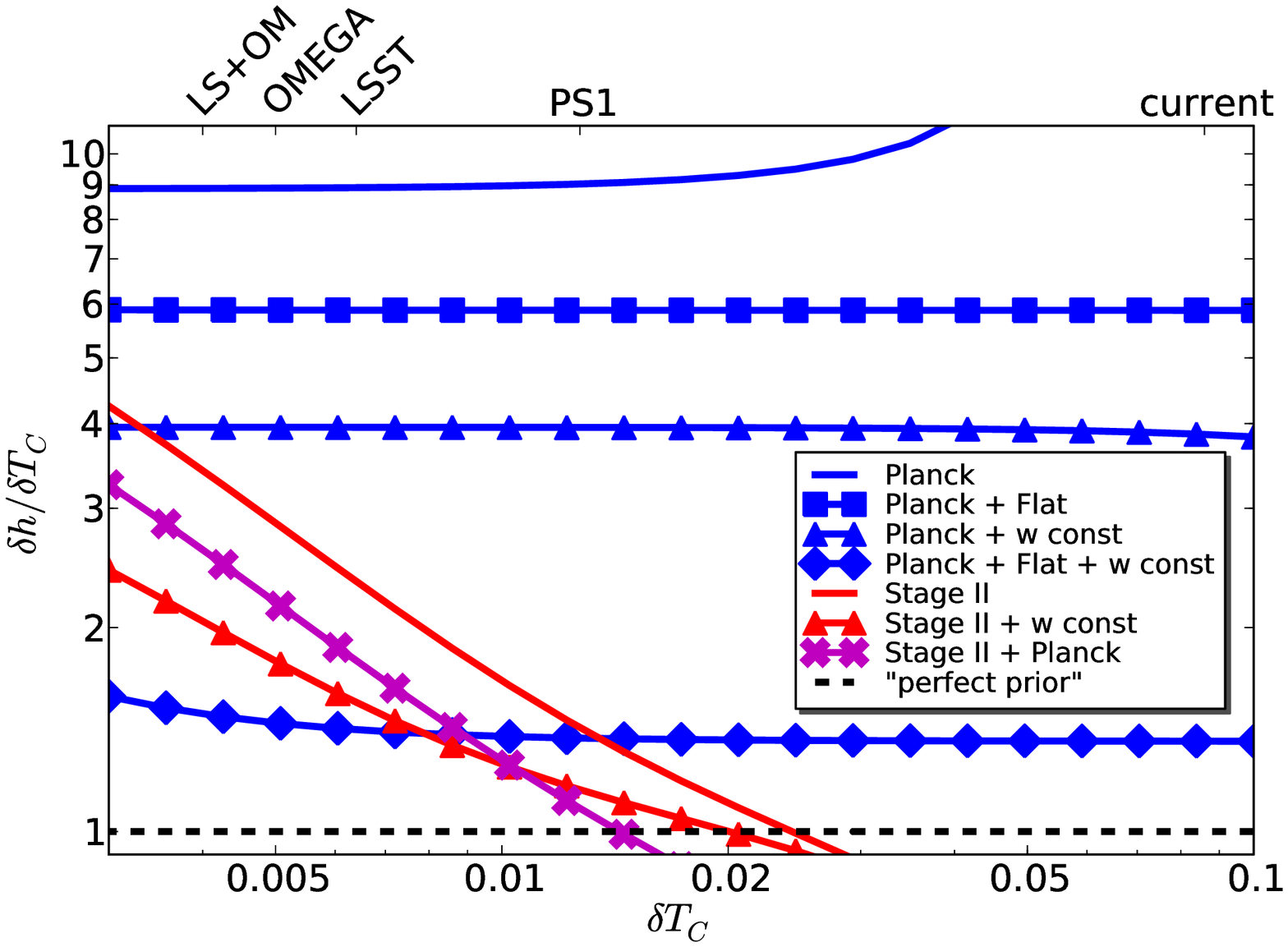}
\caption[hpriors]{
\label{fig:hpriors}%
Demonstration that $\dh > \dTC$ for future ensembles.
{\it Left}: Constraints on $h$ versus constraints on $\TC$ for various priors.
Along the top horizontal axis we plot experiments with corresponding $\dTC$:
current constraints (8.6\%), Pan-STARRS 1 (1.27\%), LSST (0.64\%), OMEGA (0.5\%), 
and LSST + OMEGA (0.4\%).
The priors are different combinations of the following:
Planck, a flat universe, constant $\w$, 
and a ``Stage II'' prior from (WL+SN+CL).
This Stage II prior constrains $\Ok$ to 0.01,
so the additional prior of flatness helps it little here.
The bottom line is the ``perfect prior'',
perfect knowledge of ($\Ode, \Om, \Ok, \wo, \wa$)
as is generally assumed,
for which $\dTC = \dh$.
{\it Right}: Relative constraints on $h$ compared to the perfect prior.
For example, given the Stage II prior, we find $\dh \sim 2.2 \dTC$.
}
\end{figure*}

This is demonstrated in Fig.~\ref{fig:hpriors} \textit{left}.
The dashed line shows $\dh = \dTC$,
or the ``perfect prior'' on ($\Om, \Ode, \Ok, \wo, \wa$)
generally assumed in analyses.
For future samples (at the left side of the plot),
as this prior is loosened, we find $\dh > \dTC$.
In Fig.~\ref{fig:hpriors} \textit{right},
we plot $\dh / \dTC$.
For example, given a ``Stage II'' prior on WL+SN+CL,
and LSST constraints on time delays ($\dTC = 0.64\%$),
we find $\dh \sim 2.2 \dTC \sim 1.4\%$.
Alternatively, assuming a Planck prior in a flat universe with constant $\w$,
we would find $\dh \sim 1.4 \dTC \sim 0.90\%$.

(Note that the Stage II WL+SN+CL prior
claims a constraint of $\dh = 0.03$,
such that it outperforms current constraints from time delays $\dh = \dTC$.)

\subsubsection{Time delays provide more than constraints on $h$}
\label{sec:morethanh}

In the introduction we commented on
the ability of any experiment to improve constraints on $\w$ and $\Ok$
simply by tightening the constraints on $h$.
Several methods have the potential to further improve the constraints on $h$ \citep{Olling07}.
Do time delays offer more than a simple constraint on $h$
for the purposes of constraining the dark energy equation of state?

\begin{figure*}
\plottwo{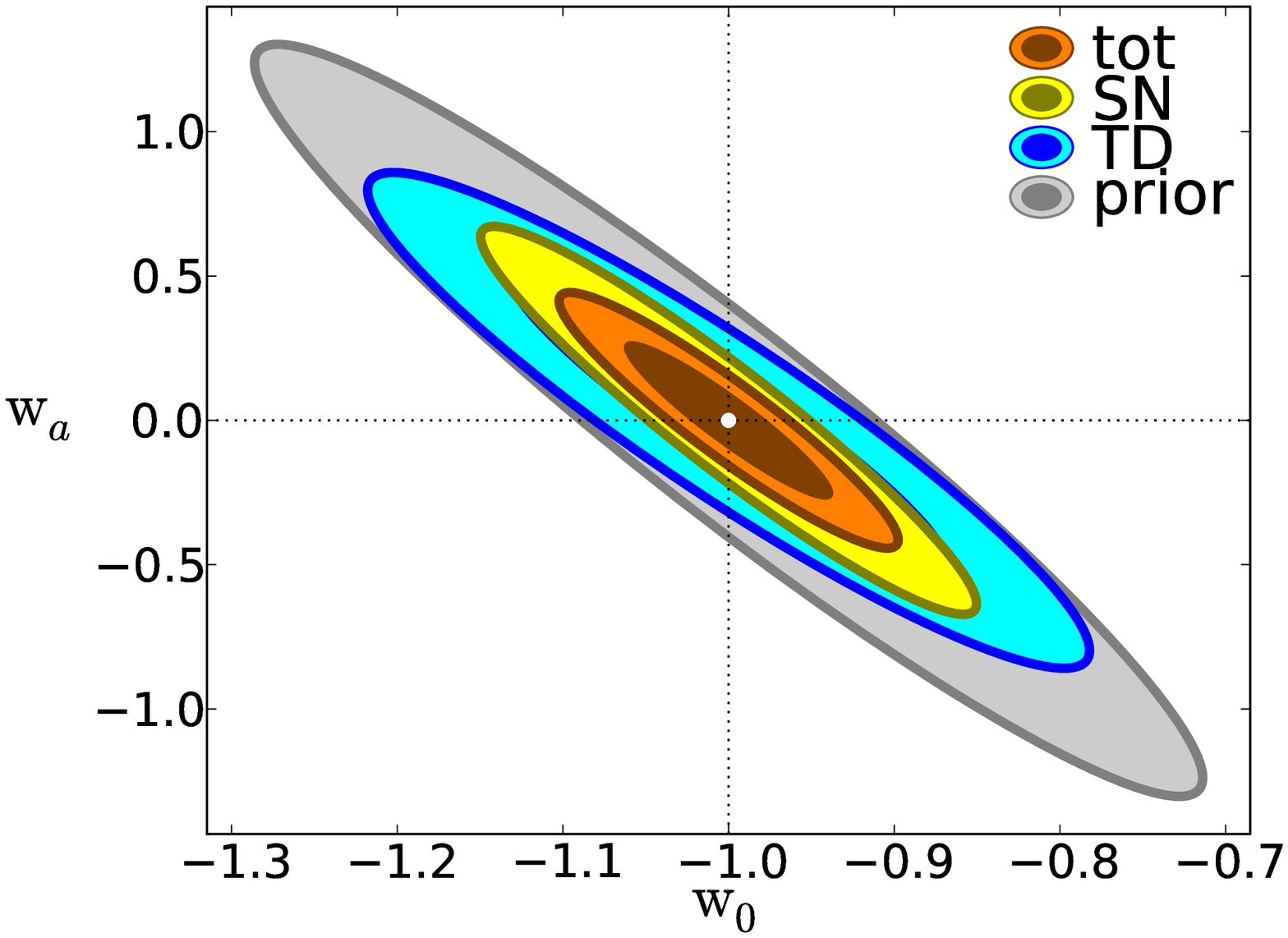}{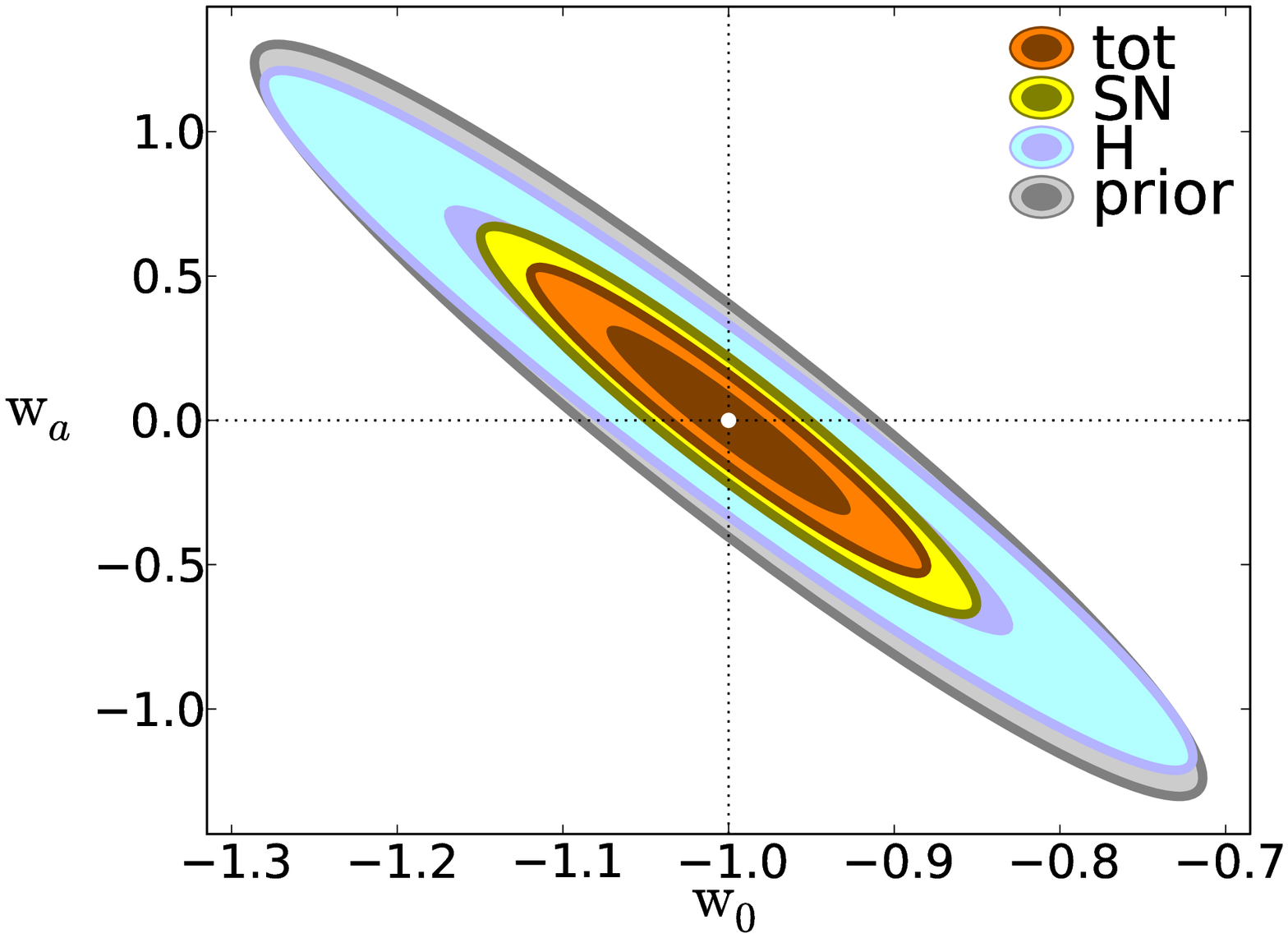}
\caption[TDvH]{
\label{fig:tdvh}%
\textit{Left}: Combined constraints on ($\wo, \wa$) 
from Stage IV time delays (TD) and supernovae (SN).
A prior of Planck + Stage II (WL+SN+CL) is assumed.
The TD + prior constraint yields $\dh = 0.008$ (not shown).
\textit{Right}: Similar plot combining Stage IV SN 
with a $\dh = 0.009$ constraint on Hubble's constant
(that which also yields $\dh = 0.008$ when combined with the prior).
Time delays yield a 19\% improvement in figure of merit
(${\rm FOM} \propto \left((\wo,\wa) ~ {\rm  Ellipse ~ Area}\right)^{-1}$), 
versus the constraint on $h$ alone.
SN + TD shows the most dramatic such improvement vs.~SN + H.
Replacing SN with the other experiments (BAO, WL, CL) 
we find lesser improvements vs.~H of
7\%, 5\%, and 3\%, respectively.
}
\end{figure*}

In Fig.~\ref{fig:tdvh} we compare
Stage IV time delays ({\it left}) to a simple $h$ constraint ({\it right})
in ability to constrain dark energy.
Each is combined with Stage IV supernovae constraints
plus a prior of Planck + Stage II WL+SN+CL.\footnote{Strictly speaking
we have not taken the proper care in combining constraints 
from the Stage II and Stage IV supernova experiments,
as their nuisance parameters have been marginalized over in the DETF Fisher matrices.
But this analysis will suffice for illustrative purposes here.}
We find time delays are more powerful than the simple $h$ constraint.
The (SN + TD + prior) figure of merit (FOM) on ($\wo, \wa$)
is $\sim 19\%$ higher than that from (SN + H + prior).

The ``H'' constraint $\dh = 0.009$ was chosen such that
when combined with the prior, the resulting $\dh$ would equal that from TD + prior.
Both H + prior and TD + prior yield $\dh = 0.008$.
However we find TD outperforms even a perfect H prior ($\dh \sim 0$) by 13\%.
Simply put, the time delay constraints on ($\Om, \Ode, \Ok, \wo, \wa$)
are clearly making contributions.

When combined with experiments other than SN,
TD offers less marked improvements over H constraints.
Replacing SN with BAO, WL, and CL,
we find TD outperforms H by 7\%, 5\%, and 3\%, respectively.

\subsection{Lens and Source Redshift Distribution}
\label{sec:zdist}

We have been considering the Gaussian redshift distributions
$z_L = 0.5 \pm 0.15$,
$z_S = 2.0 \pm 0.75$
introduced by \cite{Dobke09}
as reasonable approximate assumptions for near-future missions.
We find that the cosmological parameter constraints
are not extremely sensitive to variations in these redshift distributions.

For $\dTC = 0.64\%$ plus our Planck + Stage II (WL+SN+CL) prior,
we find the following.
A lower tighter lens redshift distribution of $z_L = 0.2 \pm 0.1$
improves the constraint on $h$ by 22\% and on $\Ode$ by 12\%
at the expense of the $\wo$ and $\wa$ constraints,
which degrade by 8\% and 10\%, respectively.
A higher tighter lens redshift distribution of $z_L = 1.0 \pm 0.1$
has less leverage,
as the $h$ and $\Ode$ constraints degrade by 15\% and 14\%, respectively
with little benefit to the other parameters.
Neither broader lens redshift distributions 
nor variations on the source redshift distribution
have much impact on the parameter constraints.

When time delay constraints are tighter ($\dTC < 0.64\%$),
with the same priors,
the lens redshift distribution begins to have a greater impact.
We reserve study of such ``beyond Stage IV'' constraints
for future work.

\section{Systematics}
\label{sec:systematics}

As with any measurement, there are many potential sources of systematic bias,
as alluded to throughout this work.
At the risk of putting the cart before the horse,
we have presented systematic-free projections for time delay cosmological constraints.
These should serve to motivate a more considered look at systematics,
in the context of the behavior of random uncertainties in these studies.
Ideally, efforts should be undertaken to reduce systematics
on a timescale comparable to that presented here
(e.g., 0.64\% by ``Stage IV'').
If this cannot be accomplished,
we study prospects for estimating cosmological parameters
in spite of large residual systematic biases in Paper III
\citep{CoeMoustakas09c}.

Here we discuss briefly the greatest potential sources of systematic bias.
We should consider which of our main sources of statistical uncertainty
(lens modelling, redshift measurements, and time delay measurements)
could also contribute significant systematic bias.
Time delay uncertainties are generally not expected to be biased in any preferred direction.
Redshift biases are somewhat worrisome but will not be discussed further here.
Most daunting are potential biases due to imprecise lens modeling.

 
Whether we determine the appropriate lens model
for the ``typical'' (``average'') lens in an ensemble
or we constrain each individual lens model well,
we must use the following tools to measure lens properties.
The largest statistical uncertainties and potential systematic biases
involve measurements of the lens mass density slope 
and perturbing mass sheets.

\subsection{Lens Mass Density Slope}
\label{sec:massslope}

Regarding mass slope, this paper has focused on the statistical strategy
which assumes that we know the correct mean
of mass slopes.
Evidence currently suggests that lenses are isothermal 
($\alpha = 1$, $\gamma = 2$)\footnote{We use two definitions common in the literature
regarding lens slope:
two-dimensional mass surface density $\kappa \propto r^{2 - \alpha}$,
and three-dimensional mass surface density $\rho \propto r^{-\gamma}$.
These parameters are related by $\alpha + \gamma \approx 3$
\citep[see discussion in][]{vandeVen09}.}
on average.
Yet a recent analysis of 58 SLACS lenses finds 
a slightly higher average slope of $\gamma = 2.085^{+0.025}_{-0.018} (\stat) \pm 0.1 (\syst)$
\citep{Koopmans09}.
If the average proved to be exactly $\gamma = 2.085$,
this would result in an 8.5\% bias in $\TC$
($\dTC = \delta \gamma / 2 = \delta \alpha$)
were we to assume an average of $\gamma = 2$ instead.

Mass profile slopes for individual lenses 
are determined by measuring mass within two radii:
the Einstein radius (from the positions of multiple images)
and a smaller radius (from velocity dispersions).
The latter require detailed spectroscopy \citep[e.g.,][]{Koopmans06}.
It will not be feasible to obtain the required measurements
for all time delay lenses detected in future surveys,
but small samples of these can be selected for such detailed study.

\subsection{Mass Sheets}
\label{sec:masssheet}

Mass sheets can be equally harmful as a source of systematics
as $\TC$ bias also scales linearly with projected mass density, 
$\dTC \sim \kappa$.
Mass sheets can result from both 
mass within the lens group environment
and mass along the line of sight (over- or under-densities) 
all the way from source to observer.
The former is the dominant effect.
Simulations \citep{DalalWatson05} suggest that group members contribute
$\kenv = 0.03 \pm 0.6$ dex
(i.e., $\log_{10}(\kenv) = \log_{10}(0.03) \pm 0.6$)
for a 1-$\sigma$ upper bound of $\kenv = 0.12$,
or 12\% bias on $\TC$.
Mass along the line of sight is generally lower
and more nearly fluctuates about the cosmic average 
but should also be accounted for.
\cite{Hilbert07} measured 
mass along the lines of sight to strong lenses in the Millennium simulation.
For sources at $z_S = 2$, the central 68\% span $-0.0355 < \klos < 0.0475$
(Paper I).

Efforts are made to measure $\kenv$ for individual lenses
via spectroscopic (and photometric) studies \citep[e.g.,][]{Momcheva06,Auger08}
and simulations which estimate the effects of nearby neighbors
\citep[e.g.,][]{KeetonZabludoff04, DalalWatson05}.
Similar studies also attempt to identify groups along the line of sight
and estimate their mass sheet contributions \citep[e.g.,][]{Fassnacht06}.

The alternative is a statistical approach.
Measurements of $\kenv$ or $\klos$ would not be required for individual lenses
if we had knowledge of the distributions $P(\kenv)$ and $P(\klos)$ for strong lenses.
These distributions could be obtained from simulations,
and one could attempt to correct for the expected bias for lenses to reside in high density regions
\citep{DalalWatson05, Oguri05b}.
However, one might wonder whether these distributions and corrections 
would prove accurate to the percent level.
Any errors would yield residual systematics in our estimation of $\TC$.

To aid such a statistical approach,
lenses in obvious groups can be excluded from the analysis
leaving only those systems with low $\kenv$.
Such low mass systems would introduce smaller biases,
though a detailed exploration of this approach will await future work.

\section{Conclusions}
\label{sec:conclusions}

We have presented the first analysis 
of the potential of gravitational lens time delays
to constrain a broad range of cosmological parameters.
The cosmological constraining power $\dTC$
was calculated for Pan-STARRS 1, LSST, and OMEGA
based on expected numbers of lenses
(including the quad-to-double ratio)
as well as the expected uncertainties in 
lens models, photometric redshifts, and time delays.
Our Fisher matrix results are provided to allow time delay constraints 
to be easily combined with and compared to constraints from other methods.

We concentrate on ``Stage IV'' constraints from LSST.
In a flat universe with constant $\w$ including a Planck prior,
LSST time delay measurements for $\sim 4,000$ lenses
should constrain 
$h$ to $\sim 0.007$ ($\sim 1\%$), 
$\Ode$ to $\sim 0.005$, and 
$\w$ to $\sim 0.026$ (all 1-$\sigma$ precisions).
We compare these results
as well as those for a general cosmology
to other ``optimistic Stage IV'' constraints expected from 
weak lensing, supernovae,  baryon acoustic oscillations, and cluster counts,
as calculated by the Dark Energy Task Force (DETF).

Combined with appropriate priors (those adopted by the DETF),
time delays provide modest constraints on a time-varying $\w(z)$
that complement the constraints expected from other methods.
Time delays constrain $\w$ best at $z \approx 0.31$,
the ``pivot redshift'' for this method.

We find that LSST and OMEGA represent about an even trade in
``quantity versus quality''
in terms of constraining cosmology with time delays.
LSST could yield $\dTC \sim 0.64\%$
by measuring time delays for 4,000 lenses,
while OMEGA could yield $\dTC \sim 0.5\%$
by obtaining high-precision time delay measurements and lens model constraints 
for 100 lenses with spectroscopic redshifts.
The combined statistical power of these two missions
could further improve the cosmological constraints to $\dTC \sim 0.4\%$.

\acknowledgements

We acknowledge useful conversations with 
Phil Marshall, Matt Auger, Chuck Keeton, Chris Kochanek, Ben Dobke,
Chris Fassnacht, Lloyd Knox, Jason Dick, Andreas Albrecht, Tony Tyson, and Jason Rhodes.
We are grateful to the DETF for releasing Fisher matrices 
detailing their estimates of cosmological constraints from various experiments.
We thank the referee for useful comments 
which led us to significantly improve the manuscript.
This work was carried out at Jet Propulsion Laboratory,
California Institute of Technology, under a contract with NASA.

\bibliographystyle{astroads}
\bibliography{paperstrunc}

\end{document}